\documentclass[
reprint,
superscriptaddress,
bibnotes,
amsmath,amssymb, aps,
nolongbibliography
]{revtex4-2}

\usepackage{graphicx} 
\usepackage{dcolumn} 
\usepackage{bm} 

\usepackage[utf8]{inputenc}
\usepackage{booktabs} 
\usepackage{float}
\usepackage{multirow}
\usepackage{color}
\usepackage[normalem]{ulem}
\usepackage{amsmath}
\usepackage{gensymb}
\usepackage{braket}
\usepackage[colorlinks,linkcolor=blue,citecolor=red,urlcolor=red]{hyperref}

\begin{document}
	\preprint{APS/123-QED}	
	\title{Large-scale quantum-dot-lithium-niobate hybrid integrated photonic circuits enabling on-chip quantum networking}
	
	\author{Xudong Wang}
	\author{Xiuqi Zhang}
	\author{Bowen Chen}
	\author{Yifan Zhu}
	\author{Yuanhao Qin}
	\author{Lvbin Dong}
	\author{Jiachen Cai}
	\author{Dongchen Sui}
	\author{Jinbo Wu}
	\author{Quan Zhang}
	\affiliation{State Key Laboratory of Materials for Integrated Circuits, Shanghai Institute of Microsystem and Information Technology, Chinese Academy of Sciences, 865 Changning Road, Shanghai, 200050, China}
	\affiliation{Center of Materials Science and Optoelectronics Engineering, University of Chinese Academy of Sciences, Beijing, 100049, China}
    \author{Runze Liu}
	\affiliation{Department of Physics, The Chinese University of Hong Kong, Hong Kong, China}
	\author{Yongheng Huo}
	\email{yongheng@ustc.edu.cn}
	\affiliation{Hefei National Research Center for Physical Sciences at the Microscale and School of Physical Sciences, University of Science and Technology of China, Hefei, 230026, China}
	\affiliation{Shanghai Research Center for Quantum Science and CAS Center for Excellence in Quantum Information and Quantum Physics, University of Science and Technology of China, Shanghai, 201315, China}
        \author{Jin Liu}
        \email{liujin23@mail.sysu.edu.cn}
        \affiliation{State Key Laboratory of Optoelectronic Materials and Technologies, School of Physics, School of Electronics and Information Technology, Sun Yat-sen University, Guangzhou 510275, China}
    \author{Xin Ou}
	\email{ouxin@mail.sim.ac.cn}
	\author{Jiaxiang Zhang}
	\email{jiaxiang.zhang@mail.sim.ac.cn}
	\affiliation{State Key Laboratory of Materials for Integrated Circuits, Shanghai Institute of Microsystem and Information Technology, Chinese Academy of Sciences, 865 Changning Road, Shanghai, 200050, China}
	\affiliation{Center of Materials Science and Optoelectronics Engineering, University of Chinese Academy of Sciences, Beijing, 100049, China}
	\date{\today}
	
	\begin{abstract}
		Hybrid integrated quantum photonics provides a promising approach to combine solid-state artificial atoms with fast reconfigurable photonic circuits, enabling scalable, chip-based quantum networks with enhanced qubit numbers and improved sources connectivity. Self-assembled quantum dots (QDs) are particularly attractive for this long-sought goal due to their ability to generate highly indistinguishable single photons with exceptional brightness and efficiency. The large-scale integration of QDs into low-loss photonic circuits could facilitate the creation of complex photonic quantum networks, enabling the generation, control, and transfer of entangled states via two-photon interference across distributed nodes. However, challenges related to the limited scalability in integration and significant inhomogeneous spectral broadening of QD emissions, and realization of quantum interference among independent sources have hindered progress toward this goal. Here, we present a hybrid photonic architecture that integrates arrays of QD-containing waveguides with low-loss lithium niobate (LN) photonic circuits, incorporating 20 deterministic single-photon sources (SPSs). By leveraging the piezoelectric properties of thin-film lithium niobate (TFLN), we introduce a circuit-compatible strain-tuning technique, achieving on-chip local spectral tuning of individual QD emissions by up to 7.7 meV—three orders of magnitude greater than the transform-limited linewidth of QD single-photon emission. This approach allows for the demonstration of on-chip quantum interference, achieving a visibility of 0.73 between two spatially separated  waveguide-coupled QD SPSs which are connected through 0.48 mm-long waveguides, thereby establishing a functional quantum network on the hybrid platform. The large-scale integration of spectrally tunable and quantum-interconnected QD-based SPSs into low-loss  LN photonic circuits, combined with the potential for fast electro-optical switching, paves the way for realizing compact, lightweight, and scalable quantum networks on a photonic chip.		
	\end{abstract}
	
	\keywords{Hybrid integration, self-assembled quantum dots, single-photon sources}
	\maketitle
	
Integrated quantum photonics (IQPs) have seen rapid development in recent years, and this burgeoning field has become an enabling technique for the advancement of optical quantum technologies such as quantum-enhanced communications, sensing, and computing\cite{obrien2009a,madsen2022a,Luo2023}. Enhancing the scalability of both the quantum bits (qubits) number and photonic circuit complexity for implementing  on-chip quantum networks is on the current subject of intensive research. Such networks would enable individual qubits to be interconnected using optical links, facilitating the generation, control, and distribution of entangled states across spatially separated quantum nodes. Thanks to the maturity of CMOS-compatible fabrication technologies, silicon photonics has emerged as a leading platform for IQPs, as highlighted by recent demonstrations of multi-photon entanglement and cluster states on compact photonic chips\cite{Bao2023,Wang2018}. However, their low single-photon generation probability, primarily due to the probabilistic quantum nature of the nonlinear single-photon sources based on down-conversion and four-wave mixing\cite{Silverstone2016,Wang2019}, presents the dominant impediment for the scalable development of quantum networks with large numbers of SPSs.

\begin{figure*}[bt]
\includegraphics{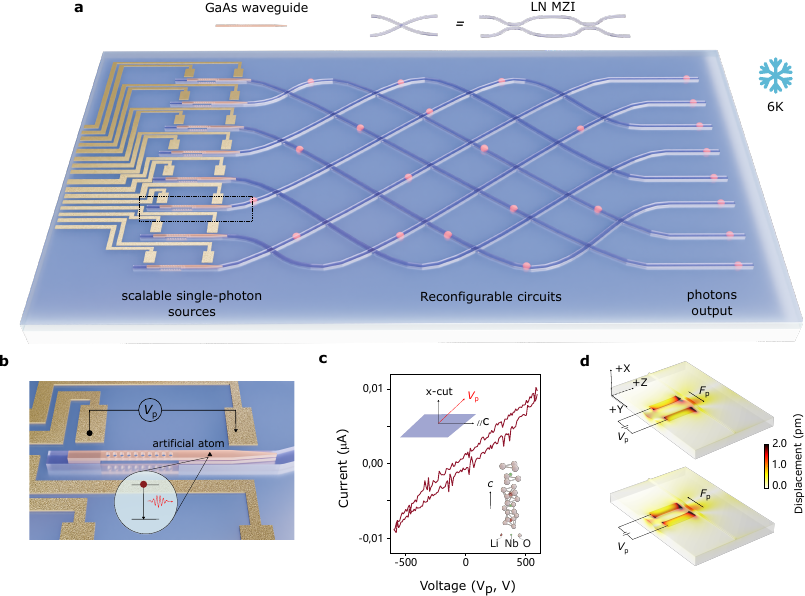}
\caption{\textbf{Large-scale chip-based quantum network on a hybrid photonic platform}. \textbf{a}, Sketch of the hybrid integrated quantum photonic chip based on III-V and LN materials.  The platform encompasses multi-channel deterministic SPSs, reconfigurable photonic mesh for qubits routing and photon output interfaces. The whole chip is operated at a cryogenic temperature. \textbf{b}, A detailed view of the core part of the hybrid photonic chip, \textit{i.e.}, GaAs/LN hybrid waveguide. It is constructed by using a step-by-step transfer-printing technique with which a GaAs waveguide is integrated in LN photonic circuits with a high alignment precision. The GaAs waveguide contains a suspended PhC nanobeam reflector and taper structure for efficient light routing from the top GaAs layer to the bottom LN waveguides.  \textbf{c}, Inverse piezoelectric effect related hysteresis of the X-cut TFLN measured at low temperature (6K). The insets are ferroelectric phase and the orientation of the applied voltage (or electric field) with respect to the \textit{c} axis of the X-cut TFLN. \textbf{d}, Simulated electric-field-induced displacements on the X-cut TFLN. The reverse displacements for $F_p$ with opposite directions indicate anisotropic strain fields.}\label{fig1:device sketch}
\end{figure*}

The pursuit of large-scale integrated quantum photonic circuits has spurred the development of hybrid integrated quantum photonics, an approach able to combine diverse functional devices from different materials into a single-chip unit. In recent years, significant efforts have been devoted to the hybrid integration of solid-state quantum emitters to enable efficient nonclassical light generation within optically-inactive photonic circuits.  Self-assembled QDs are among the most promising candidates for this goal due to their exceptional ability to generate deterministic single photons with near-unity quantum efficiency\cite{Senellart2017,Wangsps2019}. To date, QD-based SPSs have been integrated onto various photonic platforms, including silicon \cite{Kim2017, Larocque2024}, silicon nitride\cite{Davanco2017,Chanana2022}, silicon carbide \cite{Zhu2022,zhu2024}, TFLN \cite{Aghaeimeibodi2018}. Despite these advancements, a critical challenge related to the inhomogeneous spectral broadening of self-assembled QDs hinders their scalable application of QD-based SPSs across all platforms. This broadening mainly arises from random variations in shape and composition during epitaxial growth\cite{Grim2019, Patel2010}, imposing a significant obstacle for large-scale quantum applications in which qubits initialized in identical states and entanglement between remote nodes mediated by indistinguishable photons are essentially demanded. 

Thus far, numerous \textit{post}-growth tuning techniques have been explored to overcome the inhomogeneous spread in transition frequencies for chip-integrated QD-based SPSs, such as strain fields based on bulk piezoelectric substrate\cite{Jin2022,Elshaari2018,Tao2020} and thermal-optical tuning\cite{Kim2018,Katsumi2018,Osada2019,Katsumi2019,Katsumi2020}. However,  these methods are either global or hampered by high power consumption. Although local engineering of QD emission on monolithic GaAs photonic chips is possible by using the electric-field-induced quantum Stark effect\cite{Papon2023} and laser-driven local strain tuning method\cite{Grim2019}, GaAs is susceptible to high propagation loss and thus not well suited for large-scale integrated photonics. Recently, capacitive actuators have been designed for controlling the optical emission of waveguide-coupled color centers on AlN and SiN platforms\cite{Wan2020,Li2024}, enabling the integration of hundreds and thousands of color-center-based quantum emitters on hybrid photonic chips. However, this method is limited to a very small tuning range ($\sim$100 GHz), much less than the millielectronvolt scale range of the inhomogeneous broadening of self-assembled QDs\cite{Grim2019,Patel2010}. An alternative method using the lateral quantum Stark effect has been proposed for local tuning of waveguide-coupled QD emission, but the primary impediments to establishing scalable integrated quantum photonics arise from the very small tuning range of about tens of GHz as well as the accompanying large broadening of single-photon emission\cite{Larocque2024}. In this context, individual spectral control of large-scale QD-based SPSs with desired quantum connectivity via two-photon interference remains an elusive task on hybrid quantum photonic platforms.

\begin{figure*}[ht]
\includegraphics[width=1.0\textwidth]{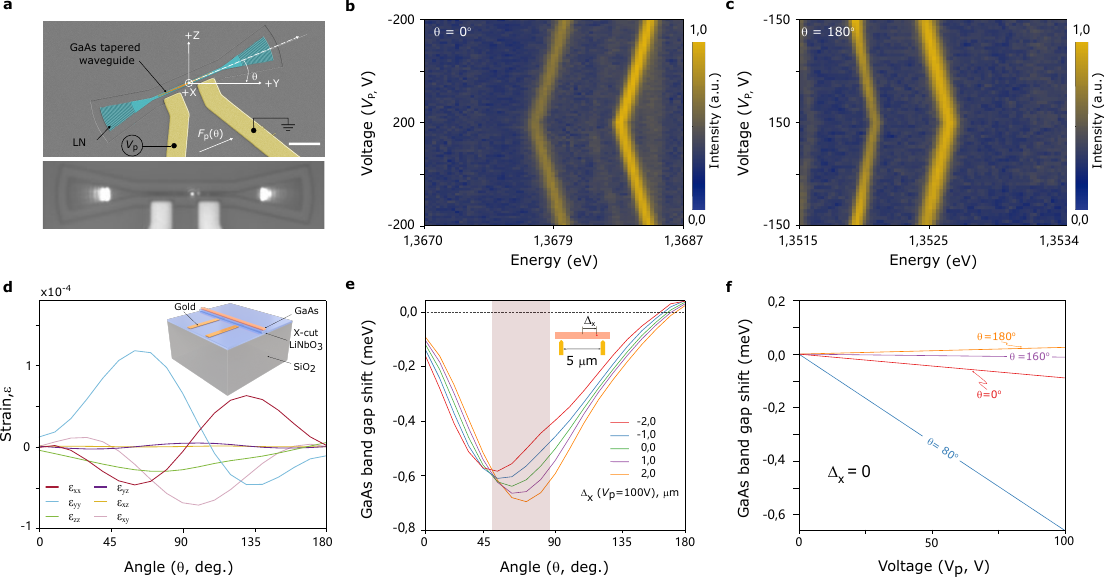}
\caption{\textbf{Anisotropic strain tuning of QD emission in hybrid III-V/LN waveguide}. \textbf{a}, Scanning electron microscopy (SEM) image (upper panel) of an representative hybrid III-V/LN waveguide. The designed structure includes a LN ridge waveguide (bright blue) with an orientation angle ($\theta$) relative to the +Y axis of the X-cut LN, the GaAs tapered waveguides (orange) bonded to LN with Van der Waals forces, and metallic electrodes (yellow).The scale bar is 10 $\mu$m. The lower panel shows a superimposed optical image of the hybrid III-V/LN waveguide and QDs photoluminescence emission coupled through the waveguide or the grating couplers. \textbf{b} and \textbf{c}, Photoluminescence spectra of QDs in the hybrid waveguide with different orientations, for $\theta=0^{\circ}$ and for $\theta=180^{\circ}$, respectively. \textbf{d}, The simulated strain field tensor for the overlaid GaAs waveguide as a function of $\theta$. In all simulations, the voltage $V_p$ is fixed at 100$V$.  The inset shows a detailed structural geometry of the III-V/LN hybrid waveguide for theoretical simulations. \textbf{e}, $\theta$ dependent GaAs band gap change for a given voltage $V_p$ at 100$V$. As the monitoring position changes away from the center of the waveguide ($\Delta_x$), the critical point for the maximum band gap change, as well as the magnitude, slightly varies. \textbf{f}, Voltage dependent change of the maximum GaAs band gap shift at the center position of the waveguide ($\Delta_x$=0). 
}\label{fig2:characterization}
\end{figure*}
    
Here we demonstrate a fully integrated and scalable hybrid quantum photonic chip for implementing quantum networking among chip-integrated solid-state quantum emitters. The hybrid platform is designed and fabricated by employing a step-by-step, high-precision  transfer-printing technique, which enables integration of waveguide-coupled QDs into LN photonic circuits with a near-unity device yield. By exploring the piezoelectric properties of the TFLN, we introduce a circuit-compatible, dynamic, and reversible local strain tuning technique to address the critical issue associated with the spectral broadening of chip-integrated QDs single-photon emission. With this technique, we achieve a large local spectral tuning range up to 7.7 meV, which is three orders of magnitude larger than the transform-limited line width of QDs emission. This allows us to demonstrate a hybrid integrated quantum photonic chip (HIQPC) that hosts more than 20 deterministic SPSs with independent spectral tuning capability. Combining the above features, we realized the scalability of chip-scale QDs-based SPSs by demonstrating a photonic quantum network via two-photon interference between two remote waveguide-coupled QDs-based SPSs that are connected through 0.48 mm-long waveguides.
	
\subsection*{Design and operational principles}
	
Fig. \ref{fig1:device sketch}a provides a schematic overview of our HIQPC, which comprises modules for deterministic single-photon generation, routing, networking, and output. An array of GaAs nanophotonic waveguides, embedded with self-assembled QDs, is integrated heterogeneously into low-loss TFLN photonic circuits using a step-by-step transfer-printing technique. The III-V/LN waveguides are bonded via van der Waals forces. Both the GaAs and LN waveguides are designed to support the fundamental TE-like mode. To ensure efficient single-photon routing from the upper GaAs waveguide to the underlying TFLN waveguide, a taper and a suspended photonic crystal (PhC) nanobeam reflector are fabricated at each end of the GaAs waveguide (Supplementary note 1). Single photons generated by QDs are evanescently coupled into a cascaded Mach-Zehnder mesh, enabling arbitrary unitary qubit manipulations, such as two-qubit entangling gates. By combining the most efficient single-photon sources—self-assembled QDs—with the highly versatile TFLN photonics platform, which exhibits strong electro-optical effects, our design facilitates the realization of advanced, chip-based optical quantum technologies with ultra-fast qubit manipulation and exceptionally low power consumption. 
\begin{figure}[bt]
\includegraphics[width=0.48\textwidth]{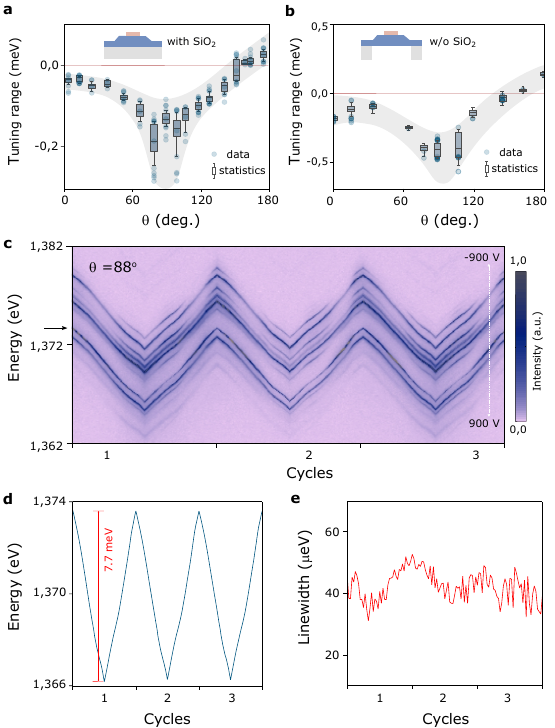}
\caption{\textbf{Statistics of strain dependent energy change of QD emission in hybrid III-V/LN waveguides}. \textbf{a} and \textbf{b}, $\theta$ dependent energy shifts of QDs in hybrid III-V/LN waveguides with and without SiO$_2$ insulating layer, respectively. Both sets of data were recorded for sweeping \textit{V}$_p$ in a fixed range from 0 to 100 V. The filled gray regions are guide to the eye. \textbf{c}, PL map of QD emission as a function of the swept voltage ($V_p$) from -900 to 900 V in the suspended device with $\theta$= 88$^\circ$. \textbf{d} and \textbf{e}, The extracted peak energy  and line width of the QD emission, which is marked by the black arrow in \textbf{c}. }\label{fig3:Performance}
\end{figure}
    
Of particular significance is that TFLN exhibits a ferroelectric phase below the Curie temperature (see inset in Fig. \ref{fig1:device sketch}c), which endows it with favorable piezoelectric properties for our hybrid integrated quantum architecture. Additionally, the piezoelectric coefficients of TFLN are temperature-insensitive, with the magnitude comparable to that of commercial lead magnesium niobate-lead titanate (PMN-PT) at cryogenic temperatures\cite{Bukhari2014,islam2019}. Inspired by these properties, here we propose a fully integrated, circuit-compatible, cryogenic strain-tuning method to control waveguide-coupled QDs emission. This is achieved by patterning electrodes close to the hybrid III-V/LN waveguide on an X-cut TFLN (see Fig. \ref{fig1:device sketch}b). When a voltage (\textit{V}$_p$, with associated electric field $\mathcal{F}_{\mathrm{p}}$) was applied along the +Y direction across the electrode with a gap size of 5 $\mu$m, a hysteresis loop was recorded as shown in Fig. \ref{fig1:device sketch}c. This indicates the presence of strain fields generated in the TFLN. The further theoretical simulations show that this strain state can be reversed as the electric field rotates by 180$^\circ$ (Fig. \ref{fig1:device sketch}d). Such anisotropy can be attributed to the crystal orientation dependence of the piezoelectric coefficients in X-cut TFLN. The applied electric field is confined to a small region, generating a localized strain field near the hybrid waveguide. This unique capability introduces additional degrees of freedom for applying independent strain fields, enabling the scalable integration of waveguide-coupled QDs, as will be discussed in later sections.
	
\subsection*{Chip-based anisotropic local strain tuning}

To explore the anisotropic strain tuning mechanism in our proposed hybrid integrated quantum photonic architecture, we fabricated a hybrid III-V/LN waveguide using the transfer printing technique (see fabrication details in Supplementary note 2). Fig. \ref{fig2:characterization}a illustrates the device structure (upper panel), where the LN waveguide is terminated by two grating couplers with an out-coupling efficiency of approximately 52.3\% at the QD emission wavelength of around 920 nm. Both ends of the GaAs waveguide are tapered from 300 nm to 50 nm over a 10 $\mu$m length, ensuring near-unity mode coupling efficiency from the upper GaAs to the underlying LN ridge waveguide. For a QD positioned at the center of the GaAs waveguide, the $\beta$-factor for the QD emission to the fundamental waveguide mode is calculated to be 83.2\%. Electrodes with a 5 $\mu$m gap is fabricated adjacent to the hybrid waveguide to apply local strain fields. As the electric field (\textit{F}$_p$) is varied, photon emission from QDs can be detected either from the top of the waveguides or the grating coupler, as shown in the lower panel of Fig.\ref{fig2:characterization}a. Upon optical pumping of a randomly selected QD in the waveguide, the generated single-photon emission can then be routed through the taper, and finally scattered to free space by the LN grating coupler. Autocorrelation measurement on the collected photons reveals a deterministic, high-purity (99.3$\pm$0.4 \%) single-photon emission (see Supplementary Fig. 1). Fig.\ref{fig2:characterization}b and \ref{fig2:characterization}c demonstrate the dynamic tuning behavior of waveguide-coupled QD emission as the voltage (V$_p$) is applied along the +Y ($\theta = 0^\circ$) and -Y ($\theta = 180^\circ$) directions, respectively. Unless otherwise specified, the photoluminescence (PL) spectra were collected from the grating coupler. Increasing \textit{V}$_p$($\theta = 0^\circ$) from -200 V to 200 V results in a redshift of $\Delta\textit{E} (\textit{E}_1-\textit{E}_0) = -0.184$ meV, while a blueshift of $\Delta$\textit{E} = 0.163 meV is observed when \textit{V}$_p$($\theta = 180^\circ$) is varied from -150 V to 150 V. The corresponding tuning rates are -0.46 $\mu$eV/V and 0.53 $\mu$eV/V, respectively. In both cases, the QD emission energy exhibits a linear dependence on V$_p$, with no quadratic dependence observed, ruling out the influence of the lateral quantum Stark effect in our device. This interpretation is further supported by a comparative study on a silica substrate with the same electrode geometry (see Supplementary Fig. 3). Additionally, no degradation in the intensity or linewidth of the QD emission is observed during the tuning process, indicating stable and precise tuning enabled by the LN piezoelectric material.

\begin{figure*}[ht]
\includegraphics[width=0.95\textwidth]{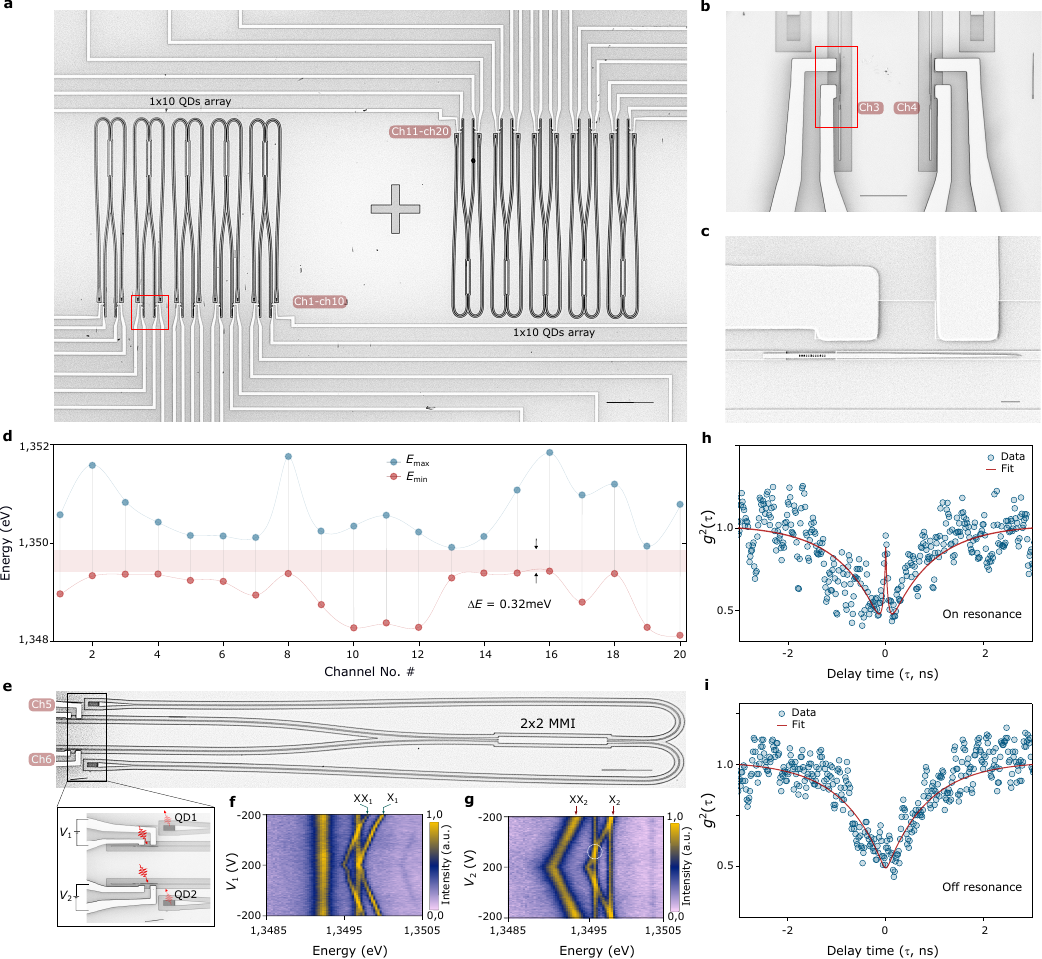}
\caption{\textbf{Hybrid integration of scalable spectrally-tunable III-V SPSs for quantum networking.} \textbf{a}, Microscopy image of a fully processed hybrid integrated quantum photonic chip based on III-V and LN materials. The hybrid platform contains two arrays of photonic modules (20 in total) and each of them is composed of III-V/LN hybrid waveguides, a MMI-based photonic routing interface and grating couplers for photons output. The scale bar is 100 $\mu$m. \textbf{b}, The zoomed-in view of the hybrid III-V/LN waveguides and the layouts of electrodes. The scale bar is 10 $\mu$m. \textbf{c}, SEM image of the hybrid III-V/LN waveguide. The scale bar is 2 $\mu$m. \textbf{d}, Independent spectral tuning of the 20 photonic quantum channels in the HIQPC. \textbf{e},  Two waveguide-coupled QDs (QD$_1$ and QD$_2$) are integrated at the two inputs of an MMI-based photonic link.  The scale bar is 50 $\mu$m. \textbf{f} and \textbf{g}, Local strain tuning of QD$_1$ and QD$_2$. X and XX refer to the exciton and biexciton photons from these two QDs. \textbf{h} and \textbf{i}, Cross-correlation of photon emission from QD$_1$ and QD$_2$ for both on- and off-resonant configurations, coupled out through one of the grating couplers at the output ports of the MMI device.}\label{fig4:large-scale integration}
\end{figure*}

Theoretically, the strain tensor ($\boldsymbol{\epsilon}$) in a QDs-containing GaAs waveguide can be written as: 
\begin{equation}
\label{eq1}
\epsilon = -s_{GaAs} e^T_{X-LN} F_p
\end{equation}
where $\boldsymbol{s}_{GaAs}$ is the compliance matrix of GaAs and $\boldsymbol{e}_{X-LN}$ is the piezoelectric constant of the X-cut LN. $\boldsymbol{e}_{X-LN}$ = $A \boldsymbol{e}_{c-axis} (N^{-1})^T$, with$A$,  $N$ and $\boldsymbol{e}_{c-axis}$ being the passive rotation matrix,  the bond strain tensor transformation matrix \cite{newnham2004} and the piezoelectric constant of the c-axis LN\cite{tarumi2012}. The full components of the strain tensor for a given voltage ($V_p$= 100 V) are calculated and visualized in Fig. \ref{fig2:characterization}d.  It shows that all strain components except for the negligible shear components ($\epsilon_{yz}$, $\epsilon_{xz}$) change significantly as the hybrid waveguide rotates about the X axis.  Plugging these quantities in Pikus-Bir Hamiltonian\cite{Sun2010} and neglecting the spin-orbit interaction, the band gap of GaAs changes with strain as follows:

\begin{equation}
\label{eq2:badgap}
\Delta E_{GaAs} = (a_c+a_v)\epsilon_{h} - \sqrt{|Q_{\epsilon}|^2+|R_{\epsilon}|^2}
\end{equation}
where $\epsilon_h = \epsilon_{xx}+ \epsilon_{yy}+ \epsilon_{zz}$ is the hydrostatic strain, $Q_{\epsilon}=-b/2(\epsilon_{xx}+ \epsilon_{yy}-2\epsilon_{zz})$, $R_{\epsilon}=\sqrt{3}/2 b(\epsilon_{xx}-\epsilon_{yy})-id\epsilon_{xy}$, and $a_c$, $a_v$, $b$, $d$ are the deformation potentials of GaAs (see Supplementary Note 3 and Supplementary Table 1). 

Fig. \ref{fig2:characterization}e and \ref{fig2:characterization}f summarize the strain-dependent change of the GaAs band gap. For a wide range of angles (0$^\circ$$\sim$160$^\circ$), applying a positive voltage leads to a GaAs band gap reduction, thus inducing a redshift in the QD emission energy. Reverse tuning occurs at angles above 160$^\circ$, a result that reproduces well the above experimental observations.  In addition, the maximum value of the energy shift can be achieved at about 70$^\circ$, and shifts slightly between 50$^\circ$ to 85$^\circ$ depending on the position away from the center of the electrode ($\Delta_x$). This is probably due to the inhomogeneous electric field distribution with the current electrode layout.  More interestingly,  there exists a critical point ($\theta\sim$160$^\circ$) at which the band gap change is insensitive to the externally applied voltage (see Supplementary Fig. 4).  

To validate the theoretical calculations, we performed statistical measurements on hybrid III-V/LN waveguides with varying orientations. As shown in Fig. \ref{fig3:Performance}a, the maximum shift in QD emission reaches 0.25 meV (with \textit{V}$_p$ ranging from 0 to 200 V) within the angle range of 75$^\circ$ to 100$^\circ$. The energy shift is smaller than the theoretical predictions in Fig. \ref{fig2:characterization}e, likely due to the strain loss uncertainties at the GaAs-LN interface formed by van der Waals forces. To enhance the device performance, we suspend the hybrid waveguide by undercutting the bottom 4.7 $\mu$m thick SiO$_2$ layer, a step designed to eliminate clamping force from the top hybrid waveguide. Fig. \ref{fig3:Performance}b demonstrates the improved performance of the device, with a maximum energy shift of 0.54 meV--double that of the previous device with the SiO$_2$ layer. By fixing the angle at the optimal value of 88°, we investigated the maximum tunability of the device. Sweeping \textit{V}$_p$ from -900 to 900 $V$, we achieved a remarkable spectral tuning range of up to 7.7 meV in the suspended device (see Fig. \ref{fig3:Performance}c). This represents the largest tuning range reported to date in a hybrid integrated quantum photonic architecture and is three orders of magnitude greater than the lifetime-limited linewidth of QD single-photon emission \cite{Kuhlmann2015,Thyrrestrup2018}.  Notably, this tuning range is comparable to that of bulky strain-tunable devices employing specialized piezoelectric actuators \cite{Zhang2013,Trotta2012}. Repeated cycling of \textit{V}$_p$ confirmed the reproducibility of the maximum spectral tuning range (Fig. \ref{fig3:Performance}d), with no significant change in the QD emission linewidth observed (Fig. \ref{fig3:Performance}e),  suggesting the excellent repeatability and stability of our method.  

\subsection*{Large-scale integration of spectrally tunable SPSs for quantum networking}
Leveraging advanced strain engineering capabilities, we have successfully fabricated a III-V/LN HIQPC, as outlined in Fig. \ref{fig1:device sketch}a. The fabrication process began with separately preparing GaAs tapered waveguides and LN photonic circuits, including multimode interferometer (MMI) beam splitter arrays, on an as-grown QD wafer and an X-cut LNOI wafer, respectively. Using a transfer printing technique, we picked up the GaAs nanophotonic waveguides from the processed QD substrate with a transparent polydimethylsiloxane (PDMS) stamp. Under high-magnification optical microscopy, alignment accuracy within hundreds of nanometers was achieved during the assembly of the HIQPC. Due to the pristine surfaces of the processed samples, the GaAs waveguides bonded tightly to the LN waveguides, even when subjected to minor disturbances from the PDMS during multiple transfer-printing steps. Through a step-by-step transfer printing process, we successfully realized a III-V/LN HIQPC integrating 20 quantum channels, each containing QD-based SPSs. This method can be further scaled to integrate an even larger number of devices if required. Fig. \ref{fig4:large-scale integration}a provides an optical microscope image of the fabricated chip, displaying two arrays of MMI beam splitters. A detailed view of the marked region is shown in Fig. \ref{fig4:large-scale integration}b, where the integrated GaAs waveguides atop the LN waveguides are clearly visible. The hybrid waveguides are aligned along the +Y axis of the X-cut TFLN, with each waveguide paired with electrodes to apply strain fields for individual control of the optical properties of the waveguide-coupled QDs. Fig. \ref{fig4:large-scale integration}c presents a SEM image of the marked region in Fig. \ref{fig4:large-scale integration}b, highlighting key features of the fabricated hybrid waveguide, including the sharp taper ends, nanobeam reflectors, and precise alignment of the two waveguides. Additionally, a complete set of optical microscopic images of the photonic quantum channels is provided in Supplementary Fig. 5, confirming a 100\% device yield achieved through the modified transfer printing technique.

We now consider a internode source connection on this hybrid platform for a chip-based quantum network. Scalable operations between photonic  quantum nodes impose two stringent requirements on waveguide-coupled QD-based quantum emitters. First, each node must be connected \textit{via} a waveguide interface for efficient photon transmission. Second, due to the inevitable inhomogeneous broadening of self-assembled QDs, the individual nodes must be spectrally tunable to achieve coherent interactions mediated by indistinguishable single photons. While the basic building block we designed in our HIQPC includes a hybrid III-V/LN waveguide comprising tapered and reflector structures for efficient single-photon generation and routing, and a 3 dB power splitter based on a 2$\times$2 MMI for internode connections, thus constituting an efficient photonic interface for multi-node source connections. For the spectral tuning of individual nodes, we apply independent voltages to electrodes fabricated close to each quantum channel. Fig. \ref{fig4:large-scale integration}d plots the spectral tuning results for all 20 photonic quantum channels. For post-selective QDs with a center energy at 1.3496 $eV$, an average energy shift of 0.82 meV is observed for a voltage swept from -100 to 100V. A spectral overlap of about 0.32 meV is obtained, indicating that the spectral discrepancies of all photonic quantum channels can be well eliminated in our HIQPC. This capability underscores the potential for scalable and coherent quantum operations on this hybrid platform.

Next, we perform on-chip two-photon interference (TPI) between two spatially separated QDs in ch5 and ch6 respectively, interconnected via a 2$\times$2 MMI beam splitter with two 0.48mm-long input waveguides (Fig. \ref{fig4:large-scale integration}e). Leveraging the broadband 3 dB power splitting capability of the MMI, we measured the emission spectra of the QDs, denoted as QD$_1$ and QD$_2$, through one output port of the MMI, while exciting both QDs simultaneously. The spectra, shown in Figs. \ref{fig4:large-scale integration}f and g, illustrate the local strain tuning behavior. Applying voltage V$_1$ shifts the emission energy of QD$_1$ while leaving QD$_2$ unaffected, and vice versa. Once the spectra of the two QDs are tuned into resonance (indicated by the white circle), we investigate the two-photon interference of the chip-routed single-photon emissions from QD$_1$ and QD$_2$. Rather than employing a conventional Hong-Ou-Mandel (HOM) setup with two single-photon detectors at opposite outputs of the beam splitter, we utilize a fiber-based Hanbury-Brown-Twiss (HBT) setup to measure the destructive interference from one output port of the MMI\cite{kim2016a}. When identical single photons from the two QDs impinge on the MMI beam splitter, the destructive interference generates the biphoton state $\ket{\Phi}$=($\ket{2,0}$-$\ket{0,2}$)/$\sqrt{2}$, indicating that the coalesced biphoton will always exit from the same output port of the MMI. By placing the HBT setup at this output, we measure the two-photon interference through the observation of the bunching at zero delay time\cite{zhe-yu2007,stevenson2013a}. Figs. \ref{fig4:large-scale integration}h and i present the normalized coincidence histograms for indistinguishable and distinguishable photons, corresponding to the on- and off-resonant tuning of the QD emission frequencies, respectively. Theoretical fits (solid lines) reveals the second-order correlation functions \textit{g}$^{(2)}$$_{on}$(0) = 0.879$\pm$0.035 and \textit{g}$^{(2)}$$_{off}$(0) = 0.508$\pm$0.011 by taking the system response time ($\delta$t) of about 99 ps into account (Supplementary Fig. 2). These values quantify the raw two-photon interference visibility as \textit{V}= (\textit{g}$^{(2)}$$_{on}$(0) - \textit{g}$^{(2)}$$_{off}$(0) )/\textit{g}$^{(2)}$$_{off}$(0) = 0.730$\pm$0.071, confirming a successful quantum link through the 0.48 mm-long low-loss photonic waveguides. Notably, this on-chip two-QD TPI experiment was conducted without temporal post-selection or background subtraction. From the theoretical fits (see Supplementary Note 4 for the model), we extract the coherence time of the QDs which is $\tau_c$=72 ps. Despite the small value, the product of 2$\delta$t/$\tau_c$ is minimized, an essential condition for observing the high interference visibility revealed in previous experiment\cite{Patel2008}.  This high visibility of the TPI can be partially attributed to the phase-stable waveguide-based photonic circuits which provide preferential TE-like mode confinement, facilitating the capture and precise alignment of the single-photon wavepackets. This is in stark contrast to free-space TPI experiments using bulk beam splitter cubes, where great care must be taken to find the optimal alignment of the biphoton wavepackets.

\section*{Discussion}
We have realized a scalable hybrid integrated quantum photonic architecture that combines the outstanding QD-based SPSs and TFLN integrated photonics with large EO effects for chip-based quantum networking. Using the piezoelectric properties of TFLN at cryogenic temperatures, we have successfully developed a circuit-compatible, highly-dynamic, reversible, and local strain-tuning technique to address key challenges associated with the inherent spectral broadening of QDs. This unique technique enables us to achieve a millielectronvolt-scale, high spatial resolution and dynamic spectral tuning capability, with the demonstration of a record-high tuning range of 7.7 meV for waveguide-coupled QDs single-photon emission on a hybrid platform. By further employing the step-by-step transfer-printing chip integration method, we have successfully fabricated the hybrid III-V/LN integrated quantum photonic chip that hosts 20 QD-based independent photonic quantum channels. Leveraging these features, we have demonstrated source scalability by realizing two-photon interference between two independent QD-based SPSs, establishing a prototype quantum network on the hybrid photonic platform.
	
The HIQPC proposed in this study uses only GaAs tapered waveguides on the LN photonic chip. Currently, QDs are not embedded in the nanophotonic cavity, so the optical properties of single-photon emission are not optimal compared to the bulky cavity-enhanced QD-based SPSs\cite{Senellart2017,Wangsps2019}. In future work, microcavities, such as planar nanobeam cavities, can be designed and fabricated into GaAs waveguides, and simultaneous application of the local strain field generated by TFLN to QDs would enable precise spectral tuning resonant with the cavity mode, which is essential for achieving Purcell-enhanced single-photon emission on chip.  We stress that this envisaged new method would offer prospects to overcome the complications of conventional thermal-optic and gas condensation tuning methods in which complex design and high-power consumption are generally required\cite{Li2024}.  Another expected improvement in our hybrid quantum photonic chip is the possibility to harness the strong EO effect of LN to design photonic circuits with an enhanced programmability, for instance,  on-chip high-speed single-photon demultiplexing, multi-mode photonic qubits manipulation,  as demonstrated in recent works\cite{Sund2023}. Given the additional strain tuning capability, our hybrid photonic platform would combine strain and EO quantum engineering knobs on a single chip unit, thus offering abundant design space for multi-purpose photonic quantum information processors. 

In addition to self-assembled QDs, our hybrid integrated photonic structure can also be applied to other solid-state quantum emitters, such as optically addressable spin defects\cite{Lukin2019,Greentree2008}, atomically thick layered semiconductors\cite{Peyskens2019,PalaciosBerraquero2017}. Since they are all strain-active quantum materials for non-classical light generation, applying local strain tuning methods will further improve source integration, thus pursuing a higher level of scalability. Finally, our III-V/LN hybrid integrated quantum photonic chip can work at cryogenic temperatures, and the power consumed for the local strain tuning in a single quantum channel can be as low as a few microwatts. Therefore, our system offers a very flexible platform to integrate the most critical superconducting nanowire single-photon detector\cite{Lomonte2021}, which provides a promising way to achieve the desired product efficiency of the detector and the source\cite{Varnava2008}, paving the path towards fault-tolerant linear optical quantum computation.

\section*{Methods}
\subsection{Materials}
The QD samples used in this study were grown by solid-source molecular beam epitaxy. The growth began with a GaAs buffer layer on a 2-inch 500 $\mu$m-thick semi-insulating (001) GaAs substrate, followed by growth of a 200 nm thick Al$_{0.8}$Ga$_{0.2}$As sacrificial layer. Then, a 160 nm thick GaAs film containing InGaAs QDs in the middle was grown. The QDs layer was formed in the Stranski-Krastanov mode and its density varied due to the temperature gradient across the GaAs substrate. $\mu$-PL mapping was conducted in order to identify the suitable QD density (10$^8$$\sim$10$^9$cm$^{-2}$) for subsequent nanophotonic device fabrication. 

The passive photonic chip was fabricated on a LNOI provided by SHNSIT. This material consists of a 300 nm thick x-cut TFLN film, a 4.7 $\mu$m SiO$_2$ layer and a 500 $\mu$m thick silicon substrate. Ion slicing and direct wafer bonding techniques were used to fabricate this heterogeneous material. In the process, a 6-inch x-cut LN crystal was first implanted with 160 keV helium ion irradiation with a fluence of 2$\times$10$^{16}$cm$^{-2}$, and then it was bonded onto a SiO$_2$/silicon wafer. With a subsequent annealing process at 150 $^{\circ}$C for 2 hours, the LN film was detached from the donor wafer and forming a standard 500 nm thick x-cut LNOI. Finally, inductively coupled plasma etching and chemical mechanical polishing techniques were used to thin the thickness of the x-cut LNOI to the target value of 300 nm for the photonic chip design and fabrication.

\subsection{Optical measurements and characterization}

The III-V/LN HIQPC was mounted on a custom-designed printed circuit board and cooled to 6 K in a closed-cycle cryostat (Montana Fusion F2) with optical windows and electrical connections. The chip was positioned on a 3-axis cryogenic nanopositioner (Nano Precision Inc., Shanghai) offering 50 nm step precision and 6 mm travel range at cryogenic temperatures. A home-built cryogenic photoluminescence imaging system was first used to pre-characterize the waveguide-coupled QDs. The system employed two LEDs (640 nm and 940 nm) for illumination, and the QD emission together with optical images of GaAs waveguides were recorded by a high-sensitivity qCMOS CCD (ORCA-Quest, Hamamatsu). A band-pass filter with a bandwidth of 10 nm in the optical path enabled wavelength pre-selection of QD emission.

A home-built micro-photoluminescence ($\mu$-PL) setup characterized the hybrid photonic chip, where QDs were excited from the top (Supplementary Fig. 6). Both excitation and collection were performed through a high numerical aperture objective (50$\times$, NA = 0.65, Mitutoyo) and coupled into single-mode fibers (SMFs). The PL spectrum was analyzed using a high-resolution spectrometer with a charge-coupled device (spectral resolution $\sim$20 pm).

For two-photon interference measurements, we implemented a modified configuration where excitation and collection paths were separated by a dichroic mirror, which provides nearly 100\% transmission for single photons. Photon coalescence from one MMI output port was extracted via a grating coupler and spectrally filtered using a home-built, tunable grating-based filter (transmission efficiency $\sim$70\%, FWHM $\sim$0.05 nm or $\sim$0.20 nm for on- and off-resonant configurations). The filtered signal was directed to a fiber-based HBT setup equipped with superconducting nanowire single-photon detectors (SNSPDs, PHOTEC, $\sim$99 ps instrumental response time). Time-correlated measurements were performed using a PicoHarp time-correlated single-photon counting (TCSPC) system.

\normalem
\bibliography{reference}

\begin{thebibliography}{49}%
\makeatletter
\providecommand \@ifxundefined [1]{%
 \@ifx{#1\undefined}
}%
\providecommand \@ifnum [1]{%
 \ifnum #1\expandafter \@firstoftwo
 \else \expandafter \@secondoftwo
 \fi
}%
\providecommand \@ifx [1]{%
 \ifx #1\expandafter \@firstoftwo
 \else \expandafter \@secondoftwo
 \fi
}%
\providecommand \natexlab [1]{#1}%
\providecommand \enquote  [1]{``#1''}%
\providecommand \bibnamefont  [1]{#1}%
\providecommand \bibfnamefont [1]{#1}%
\providecommand \citenamefont [1]{#1}%
\providecommand \href@noop [0]{\@secondoftwo}%
\providecommand \href [0]{\begingroup \@sanitize@url \@href}%
\providecommand \@href[1]{\@@startlink{#1}\@@href}%
\providecommand \@@href[1]{\endgroup#1\@@endlink}%
\providecommand \@sanitize@url [0]{\catcode `\\12\catcode `\$12\catcode `\&12\catcode `\#12\catcode `\^12\catcode `\_12\catcode `\%12\relax}%
\providecommand \@@startlink[1]{}%
\providecommand \@@endlink[0]{}%
\providecommand \url  [0]{\begingroup\@sanitize@url \@url }%
\providecommand \@url [1]{\endgroup\@href {#1}{\urlprefix }}%
\providecommand \urlprefix  [0]{URL }%
\providecommand \Eprint [0]{\href }%
\providecommand \doibase [0]{https://doi.org/}%
\providecommand \selectlanguage [0]{\@gobble}%
\providecommand \bibinfo  [0]{\@secondoftwo}%
\providecommand \bibfield  [0]{\@secondoftwo}%
\providecommand \translation [1]{[#1]}%
\providecommand \BibitemOpen [0]{}%
\providecommand \bibitemStop [0]{}%
\providecommand \bibitemNoStop [0]{.\EOS\space}%
\providecommand \EOS [0]{\spacefactor3000\relax}%
\providecommand \BibitemShut  [1]{\csname bibitem#1\endcsname}%
\let\auto@bib@innerbib\@empty
\bibitem [{\citenamefont {O'Brien}\ \emph {et~al.}(2009)\citenamefont {O'Brien}, \citenamefont {Furusawa},\ and\ \citenamefont {Vu{\v c}kovi{\'c}}}]{obrien2009a}%
  \BibitemOpen
  \bibfield  {author} {\bibinfo {author} {\bibfnamefont {J.~L.}\ \bibnamefont {O'Brien}}, \bibinfo {author} {\bibfnamefont {A.}~\bibnamefont {Furusawa}},\ and\ \bibinfo {author} {\bibfnamefont {J.}~\bibnamefont {Vu{\v c}kovi{\'c}}},\ }\href {https://doi.org/10.1038/nphoton.2009.229} {\bibfield  {journal} {\bibinfo  {journal} {Nat. Photonics}\ }\textbf {\bibinfo {volume} {3}},\ \bibinfo {pages} {687} (\bibinfo {year} {2009})}\BibitemShut {NoStop}%
\bibitem [{\citenamefont {Madsen}\ \emph {et~al.}(2022)\citenamefont {Madsen}, \citenamefont {Laudenbach}, \citenamefont {Askarani}, \citenamefont {Rortais}, \citenamefont {Vincent}, \citenamefont {Bulmer}, \citenamefont {Miatto}, \citenamefont {Neuhaus}, \citenamefont {Helt}, \citenamefont {Collins}, \citenamefont {Lita}, \citenamefont {Gerrits}, \citenamefont {Nam}, \citenamefont {Vaidya}, \citenamefont {Menotti}, \citenamefont {Dhand}, \citenamefont {Vernon}, \citenamefont {Quesada},\ and\ \citenamefont {Lavoie}}]{madsen2022a}%
  \BibitemOpen
  \bibfield  {author} {\bibinfo {author} {\bibfnamefont {L.~S.}\ \bibnamefont {Madsen}}, \bibinfo {author} {\bibfnamefont {F.}~\bibnamefont {Laudenbach}}, \bibinfo {author} {\bibfnamefont {M.~F.}\ \bibnamefont {Askarani}}, \bibinfo {author} {\bibfnamefont {F.}~\bibnamefont {Rortais}}, \bibinfo {author} {\bibfnamefont {T.}~\bibnamefont {Vincent}}, \bibinfo {author} {\bibfnamefont {J.~F.~F.}\ \bibnamefont {Bulmer}}, \bibinfo {author} {\bibfnamefont {F.~M.}\ \bibnamefont {Miatto}}, \bibinfo {author} {\bibfnamefont {L.}~\bibnamefont {Neuhaus}}, \bibinfo {author} {\bibfnamefont {L.~G.}\ \bibnamefont {Helt}}, \bibinfo {author} {\bibfnamefont {M.~J.}\ \bibnamefont {Collins}}, \bibinfo {author} {\bibfnamefont {A.~E.}\ \bibnamefont {Lita}}, \bibinfo {author} {\bibfnamefont {T.}~\bibnamefont {Gerrits}}, \bibinfo {author} {\bibfnamefont {S.~W.}\ \bibnamefont {Nam}}, \bibinfo {author} {\bibfnamefont {V.~D.}\ \bibnamefont {Vaidya}}, \bibinfo {author} {\bibfnamefont {M.}~\bibnamefont {Menotti}}, \bibinfo {author}
  {\bibfnamefont {I.}~\bibnamefont {Dhand}}, \bibinfo {author} {\bibfnamefont {Z.}~\bibnamefont {Vernon}}, \bibinfo {author} {\bibfnamefont {N.}~\bibnamefont {Quesada}},\ and\ \bibinfo {author} {\bibfnamefont {J.}~\bibnamefont {Lavoie}},\ }\href {https://doi.org/10.1038/s41586-022-04725-x} {\bibfield  {journal} {\bibinfo  {journal} {Nature}\ }\textbf {\bibinfo {volume} {606}},\ \bibinfo {pages} {75} (\bibinfo {year} {2022})}\BibitemShut {NoStop}%
\bibitem [{\citenamefont {Luo}\ \emph {et~al.}(2023)\citenamefont {Luo}, \citenamefont {Cao}, \citenamefont {Shi}, \citenamefont {Wan}, \citenamefont {Zhang}, \citenamefont {Li}, \citenamefont {Chen}, \citenamefont {Li}, \citenamefont {Li}, \citenamefont {Wang}, \citenamefont {Sun}, \citenamefont {Karim}, \citenamefont {Cai}, \citenamefont {Kwek},\ and\ \citenamefont {Liu}}]{Luo2023}%
  \BibitemOpen
  \bibfield  {author} {\bibinfo {author} {\bibfnamefont {W.}~\bibnamefont {Luo}}, \bibinfo {author} {\bibfnamefont {L.}~\bibnamefont {Cao}}, \bibinfo {author} {\bibfnamefont {Y.}~\bibnamefont {Shi}}, \bibinfo {author} {\bibfnamefont {L.}~\bibnamefont {Wan}}, \bibinfo {author} {\bibfnamefont {H.}~\bibnamefont {Zhang}}, \bibinfo {author} {\bibfnamefont {S.}~\bibnamefont {Li}}, \bibinfo {author} {\bibfnamefont {G.}~\bibnamefont {Chen}}, \bibinfo {author} {\bibfnamefont {Y.}~\bibnamefont {Li}}, \bibinfo {author} {\bibfnamefont {S.}~\bibnamefont {Li}}, \bibinfo {author} {\bibfnamefont {Y.}~\bibnamefont {Wang}}, \bibinfo {author} {\bibfnamefont {S.}~\bibnamefont {Sun}}, \bibinfo {author} {\bibfnamefont {M.~F.}\ \bibnamefont {Karim}}, \bibinfo {author} {\bibfnamefont {H.}~\bibnamefont {Cai}}, \bibinfo {author} {\bibfnamefont {L.~C.}\ \bibnamefont {Kwek}},\ and\ \bibinfo {author} {\bibfnamefont {A.~Q.}\ \bibnamefont {Liu}},\ }\bibfield  {journal} {\bibinfo  {journal} {Light: Science and Applications}\ }\textbf
  {\bibinfo {volume} {12}},\ \href {https://doi.org/10.1038/s41377-023-01173-8} {10.1038/s41377-023-01173-8} (\bibinfo {year} {2023})\BibitemShut {NoStop}%
\bibitem [{\citenamefont {Bao}\ \emph {et~al.}(2023)\citenamefont {Bao}, \citenamefont {Fu}, \citenamefont {Pramanik}, \citenamefont {Mao}, \citenamefont {Chi}, \citenamefont {Cao}, \citenamefont {Zhai}, \citenamefont {Mao}, \citenamefont {Dai}, \citenamefont {Chen}, \citenamefont {Jia}, \citenamefont {Zhao}, \citenamefont {Zheng}, \citenamefont {Tang}, \citenamefont {Li}, \citenamefont {Luo}, \citenamefont {Wang}, \citenamefont {Yang}, \citenamefont {Peng}, \citenamefont {Liu}, \citenamefont {Dai}, \citenamefont {He}, \citenamefont {Muthali}, \citenamefont {Oxenl{\o}we}, \citenamefont {Vigliar}, \citenamefont {Paesani}, \citenamefont {Hou}, \citenamefont {Santagati}, \citenamefont {Silverstone}, \citenamefont {Laing}, \citenamefont {Thompson}, \citenamefont {O'Brien}, \citenamefont {Ding}, \citenamefont {Gong},\ and\ \citenamefont {Wang}}]{Bao2023}%
  \BibitemOpen
  \bibfield  {author} {\bibinfo {author} {\bibfnamefont {J.}~\bibnamefont {Bao}}, \bibinfo {author} {\bibfnamefont {Z.}~\bibnamefont {Fu}}, \bibinfo {author} {\bibfnamefont {T.}~\bibnamefont {Pramanik}}, \bibinfo {author} {\bibfnamefont {J.}~\bibnamefont {Mao}}, \bibinfo {author} {\bibfnamefont {Y.}~\bibnamefont {Chi}}, \bibinfo {author} {\bibfnamefont {Y.}~\bibnamefont {Cao}}, \bibinfo {author} {\bibfnamefont {C.}~\bibnamefont {Zhai}}, \bibinfo {author} {\bibfnamefont {Y.}~\bibnamefont {Mao}}, \bibinfo {author} {\bibfnamefont {T.}~\bibnamefont {Dai}}, \bibinfo {author} {\bibfnamefont {X.}~\bibnamefont {Chen}}, \bibinfo {author} {\bibfnamefont {X.}~\bibnamefont {Jia}}, \bibinfo {author} {\bibfnamefont {L.}~\bibnamefont {Zhao}}, \bibinfo {author} {\bibfnamefont {Y.}~\bibnamefont {Zheng}}, \bibinfo {author} {\bibfnamefont {B.}~\bibnamefont {Tang}}, \bibinfo {author} {\bibfnamefont {Z.}~\bibnamefont {Li}}, \bibinfo {author} {\bibfnamefont {J.}~\bibnamefont {Luo}}, \bibinfo {author} {\bibfnamefont
  {W.}~\bibnamefont {Wang}}, \bibinfo {author} {\bibfnamefont {Y.}~\bibnamefont {Yang}}, \bibinfo {author} {\bibfnamefont {Y.}~\bibnamefont {Peng}}, \bibinfo {author} {\bibfnamefont {D.}~\bibnamefont {Liu}}, \bibinfo {author} {\bibfnamefont {D.}~\bibnamefont {Dai}}, \bibinfo {author} {\bibfnamefont {Q.}~\bibnamefont {He}}, \bibinfo {author} {\bibfnamefont {A.~L.}\ \bibnamefont {Muthali}}, \bibinfo {author} {\bibfnamefont {L.~K.}\ \bibnamefont {Oxenl{\o}we}}, \bibinfo {author} {\bibfnamefont {C.}~\bibnamefont {Vigliar}}, \bibinfo {author} {\bibfnamefont {S.}~\bibnamefont {Paesani}}, \bibinfo {author} {\bibfnamefont {H.}~\bibnamefont {Hou}}, \bibinfo {author} {\bibfnamefont {R.}~\bibnamefont {Santagati}}, \bibinfo {author} {\bibfnamefont {J.~W.}\ \bibnamefont {Silverstone}}, \bibinfo {author} {\bibfnamefont {A.}~\bibnamefont {Laing}}, \bibinfo {author} {\bibfnamefont {M.~G.}\ \bibnamefont {Thompson}}, \bibinfo {author} {\bibfnamefont {J.~L.}\ \bibnamefont {O'Brien}}, \bibinfo {author} {\bibfnamefont
  {Y.}~\bibnamefont {Ding}}, \bibinfo {author} {\bibfnamefont {Q.}~\bibnamefont {Gong}},\ and\ \bibinfo {author} {\bibfnamefont {J.}~\bibnamefont {Wang}},\ }\href {https://doi.org/10.1038/s41566-023-01187-z} {\bibfield  {journal} {\bibinfo  {journal} {Nature Photonics}\ }\textbf {\bibinfo {volume} {17}},\ \bibinfo {pages} {573} (\bibinfo {year} {2023})}\BibitemShut {NoStop}%
\bibitem [{\citenamefont {Wang}\ \emph {et~al.}(2018)\citenamefont {Wang}, \citenamefont {Paesani}, \citenamefont {Ding}, \citenamefont {Santagati}, \citenamefont {Skrzypczyk}, \citenamefont {Salavrakos}, \citenamefont {Tura}, \citenamefont {Augusiak}, \citenamefont {Mančinska}, \citenamefont {Bacco}, \citenamefont {Bonneau}, \citenamefont {Silverstone}, \citenamefont {Gong}, \citenamefont {Acín}, \citenamefont {Rottwitt}, \citenamefont {Oxenløwe}, \citenamefont {O’Brien}, \citenamefont {Laing},\ and\ \citenamefont {Thompson}}]{Wang2018}%
  \BibitemOpen
  \bibfield  {author} {\bibinfo {author} {\bibfnamefont {J.}~\bibnamefont {Wang}}, \bibinfo {author} {\bibfnamefont {S.}~\bibnamefont {Paesani}}, \bibinfo {author} {\bibfnamefont {Y.}~\bibnamefont {Ding}}, \bibinfo {author} {\bibfnamefont {R.}~\bibnamefont {Santagati}}, \bibinfo {author} {\bibfnamefont {P.}~\bibnamefont {Skrzypczyk}}, \bibinfo {author} {\bibfnamefont {A.}~\bibnamefont {Salavrakos}}, \bibinfo {author} {\bibfnamefont {J.}~\bibnamefont {Tura}}, \bibinfo {author} {\bibfnamefont {R.}~\bibnamefont {Augusiak}}, \bibinfo {author} {\bibfnamefont {L.}~\bibnamefont {Mančinska}}, \bibinfo {author} {\bibfnamefont {D.}~\bibnamefont {Bacco}}, \bibinfo {author} {\bibfnamefont {D.}~\bibnamefont {Bonneau}}, \bibinfo {author} {\bibfnamefont {J.~W.}\ \bibnamefont {Silverstone}}, \bibinfo {author} {\bibfnamefont {Q.}~\bibnamefont {Gong}}, \bibinfo {author} {\bibfnamefont {A.}~\bibnamefont {Acín}}, \bibinfo {author} {\bibfnamefont {K.}~\bibnamefont {Rottwitt}}, \bibinfo {author} {\bibfnamefont {L.~K.}\
  \bibnamefont {Oxenløwe}}, \bibinfo {author} {\bibfnamefont {J.~L.}\ \bibnamefont {O’Brien}}, \bibinfo {author} {\bibfnamefont {A.}~\bibnamefont {Laing}},\ and\ \bibinfo {author} {\bibfnamefont {M.~G.}\ \bibnamefont {Thompson}},\ }\href {https://doi.org/10.1126/science.aar7053} {\bibfield  {journal} {\bibinfo  {journal} {Science}\ }\textbf {\bibinfo {volume} {360}},\ \bibinfo {pages} {285–291} (\bibinfo {year} {2018})}\BibitemShut {NoStop}%
\bibitem [{\citenamefont {Silverstone}\ \emph {et~al.}(2016)\citenamefont {Silverstone}, \citenamefont {Bonneau}, \citenamefont {O’Brien},\ and\ \citenamefont {Thompson}}]{Silverstone2016}%
  \BibitemOpen
  \bibfield  {author} {\bibinfo {author} {\bibfnamefont {J.~W.}\ \bibnamefont {Silverstone}}, \bibinfo {author} {\bibfnamefont {D.}~\bibnamefont {Bonneau}}, \bibinfo {author} {\bibfnamefont {J.~L.}\ \bibnamefont {O’Brien}},\ and\ \bibinfo {author} {\bibfnamefont {M.~G.}\ \bibnamefont {Thompson}},\ }\href {https://doi.org/10.1109/jstqe.2016.2573218} {\bibfield  {journal} {\bibinfo  {journal} {IEEE Journal of Selected Topics in Quantum Electronics}\ }\textbf {\bibinfo {volume} {22}},\ \bibinfo {pages} {390–402} (\bibinfo {year} {2016})}\BibitemShut {NoStop}%
\bibitem [{\citenamefont {Wang}\ \emph {et~al.}(2020)\citenamefont {Wang}, \citenamefont {Sciarrino}, \citenamefont {Laing},\ and\ \citenamefont {Thompson}}]{Wang2019}%
  \BibitemOpen
  \bibfield  {author} {\bibinfo {author} {\bibfnamefont {J.}~\bibnamefont {Wang}}, \bibinfo {author} {\bibfnamefont {F.}~\bibnamefont {Sciarrino}}, \bibinfo {author} {\bibfnamefont {A.}~\bibnamefont {Laing}},\ and\ \bibinfo {author} {\bibfnamefont {M.~G.}\ \bibnamefont {Thompson}},\ }\href {https://doi.org/10.1038/s41566-019-0532-1} {\bibfield  {journal} {\bibinfo  {journal} {Nature Photonics}\ }\textbf {\bibinfo {volume} {14}},\ \bibinfo {pages} {273} (\bibinfo {year} {2020})}\BibitemShut {NoStop}%
\bibitem [{\citenamefont {Senellart}\ \emph {et~al.}(2017)\citenamefont {Senellart}, \citenamefont {Solomon},\ and\ \citenamefont {White}}]{Senellart2017}%
  \BibitemOpen
  \bibfield  {author} {\bibinfo {author} {\bibfnamefont {P.}~\bibnamefont {Senellart}}, \bibinfo {author} {\bibfnamefont {G.}~\bibnamefont {Solomon}},\ and\ \bibinfo {author} {\bibfnamefont {A.}~\bibnamefont {White}},\ }\href {https://doi.org/10.1038/nnano.2017.218} {\bibfield  {journal} {\bibinfo  {journal} {Nature Nanotechnology}\ }\textbf {\bibinfo {volume} {12}},\ \bibinfo {pages} {1026–1039} (\bibinfo {year} {2017})}\BibitemShut {NoStop}%
\bibitem [{\citenamefont {Wang}\ \emph {et~al.}(2019)\citenamefont {Wang}, \citenamefont {He}, \citenamefont {Chung}, \citenamefont {Hu}, \citenamefont {Yu}, \citenamefont {Chen}, \citenamefont {Ding}, \citenamefont {Chen}, \citenamefont {Qin}, \citenamefont {Yang}, \citenamefont {Liu}, \citenamefont {Duan}, \citenamefont {Li}, \citenamefont {Gerhardt}, \citenamefont {Winkler}, \citenamefont {Jurkat}, \citenamefont {Wang}, \citenamefont {Gregersen}, \citenamefont {Huo}, \citenamefont {Dai}, \citenamefont {Yu}, \citenamefont {H\"{o}fling}, \citenamefont {Lu},\ and\ \citenamefont {Pan}}]{Wangsps2019}%
  \BibitemOpen
  \bibfield  {author} {\bibinfo {author} {\bibfnamefont {H.}~\bibnamefont {Wang}}, \bibinfo {author} {\bibfnamefont {Y.-M.}\ \bibnamefont {He}}, \bibinfo {author} {\bibfnamefont {T.-H.}\ \bibnamefont {Chung}}, \bibinfo {author} {\bibfnamefont {H.}~\bibnamefont {Hu}}, \bibinfo {author} {\bibfnamefont {Y.}~\bibnamefont {Yu}}, \bibinfo {author} {\bibfnamefont {S.}~\bibnamefont {Chen}}, \bibinfo {author} {\bibfnamefont {X.}~\bibnamefont {Ding}}, \bibinfo {author} {\bibfnamefont {M.-C.}\ \bibnamefont {Chen}}, \bibinfo {author} {\bibfnamefont {J.}~\bibnamefont {Qin}}, \bibinfo {author} {\bibfnamefont {X.}~\bibnamefont {Yang}}, \bibinfo {author} {\bibfnamefont {R.-Z.}\ \bibnamefont {Liu}}, \bibinfo {author} {\bibfnamefont {Z.-C.}\ \bibnamefont {Duan}}, \bibinfo {author} {\bibfnamefont {J.-P.}\ \bibnamefont {Li}}, \bibinfo {author} {\bibfnamefont {S.}~\bibnamefont {Gerhardt}}, \bibinfo {author} {\bibfnamefont {K.}~\bibnamefont {Winkler}}, \bibinfo {author} {\bibfnamefont {J.}~\bibnamefont {Jurkat}}, \bibinfo {author}
  {\bibfnamefont {L.-J.}\ \bibnamefont {Wang}}, \bibinfo {author} {\bibfnamefont {N.}~\bibnamefont {Gregersen}}, \bibinfo {author} {\bibfnamefont {Y.-H.}\ \bibnamefont {Huo}}, \bibinfo {author} {\bibfnamefont {Q.}~\bibnamefont {Dai}}, \bibinfo {author} {\bibfnamefont {S.}~\bibnamefont {Yu}}, \bibinfo {author} {\bibfnamefont {S.}~\bibnamefont {H\"{o}fling}}, \bibinfo {author} {\bibfnamefont {C.-Y.}\ \bibnamefont {Lu}},\ and\ \bibinfo {author} {\bibfnamefont {J.-W.}\ \bibnamefont {Pan}},\ }\href {https://doi.org/10.1038/s41566-019-0494-3} {\bibfield  {journal} {\bibinfo  {journal} {Nature Photonics}\ }\textbf {\bibinfo {volume} {13}},\ \bibinfo {pages} {770–775} (\bibinfo {year} {2019})}\BibitemShut {NoStop}%
\bibitem [{\citenamefont {Kim}\ \emph {et~al.}(2017)\citenamefont {Kim}, \citenamefont {Aghaeimeibodi}, \citenamefont {Richardson}, \citenamefont {Leavitt}, \citenamefont {Englund},\ and\ \citenamefont {Waks}}]{Kim2017}%
  \BibitemOpen
  \bibfield  {author} {\bibinfo {author} {\bibfnamefont {J.-H.}\ \bibnamefont {Kim}}, \bibinfo {author} {\bibfnamefont {S.}~\bibnamefont {Aghaeimeibodi}}, \bibinfo {author} {\bibfnamefont {C.~J.~K.}\ \bibnamefont {Richardson}}, \bibinfo {author} {\bibfnamefont {R.~P.}\ \bibnamefont {Leavitt}}, \bibinfo {author} {\bibfnamefont {D.}~\bibnamefont {Englund}},\ and\ \bibinfo {author} {\bibfnamefont {E.}~\bibnamefont {Waks}},\ }\href {https://doi.org/10.1021/acs.nanolett.7b03220} {\bibfield  {journal} {\bibinfo  {journal} {Nano Letters}\ }\textbf {\bibinfo {volume} {17}},\ \bibinfo {pages} {7394–7400} (\bibinfo {year} {2017})}\BibitemShut {NoStop}%
\bibitem [{\citenamefont {Larocque}\ \emph {et~al.}(2024)\citenamefont {Larocque}, \citenamefont {Buyukkaya}, \citenamefont {Errando-Herranz}, \citenamefont {Papon}, \citenamefont {Harper}, \citenamefont {Tao}, \citenamefont {Carolan}, \citenamefont {Lee}, \citenamefont {Richardson}, \citenamefont {Leake}, \citenamefont {Coleman}, \citenamefont {Fanto}, \citenamefont {Waks},\ and\ \citenamefont {Englund}}]{Larocque2024}%
  \BibitemOpen
  \bibfield  {author} {\bibinfo {author} {\bibfnamefont {H.}~\bibnamefont {Larocque}}, \bibinfo {author} {\bibfnamefont {M.~A.}\ \bibnamefont {Buyukkaya}}, \bibinfo {author} {\bibfnamefont {C.}~\bibnamefont {Errando-Herranz}}, \bibinfo {author} {\bibfnamefont {C.}~\bibnamefont {Papon}}, \bibinfo {author} {\bibfnamefont {S.}~\bibnamefont {Harper}}, \bibinfo {author} {\bibfnamefont {M.}~\bibnamefont {Tao}}, \bibinfo {author} {\bibfnamefont {J.}~\bibnamefont {Carolan}}, \bibinfo {author} {\bibfnamefont {C.-M.}\ \bibnamefont {Lee}}, \bibinfo {author} {\bibfnamefont {C.~J.~K.}\ \bibnamefont {Richardson}}, \bibinfo {author} {\bibfnamefont {G.~L.}\ \bibnamefont {Leake}}, \bibinfo {author} {\bibfnamefont {D.~J.}\ \bibnamefont {Coleman}}, \bibinfo {author} {\bibfnamefont {M.~L.}\ \bibnamefont {Fanto}}, \bibinfo {author} {\bibfnamefont {E.}~\bibnamefont {Waks}},\ and\ \bibinfo {author} {\bibfnamefont {D.}~\bibnamefont {Englund}},\ }\bibfield  {journal} {\bibinfo  {journal} {Nature Communications}\ }\textbf {\bibinfo
  {volume} {15}},\ \href {https://doi.org/10.1038/s41467-024-50208-0} {10.1038/s41467-024-50208-0} (\bibinfo {year} {2024})\BibitemShut {NoStop}%
\bibitem [{\citenamefont {Davanco}\ \emph {et~al.}(2017)\citenamefont {Davanco}, \citenamefont {Liu}, \citenamefont {Sapienza}, \citenamefont {Zhang}, \citenamefont {De~Miranda~Cardoso}, \citenamefont {Verma}, \citenamefont {Mirin}, \citenamefont {Nam}, \citenamefont {Liu},\ and\ \citenamefont {Srinivasan}}]{Davanco2017}%
  \BibitemOpen
  \bibfield  {author} {\bibinfo {author} {\bibfnamefont {M.}~\bibnamefont {Davanco}}, \bibinfo {author} {\bibfnamefont {J.}~\bibnamefont {Liu}}, \bibinfo {author} {\bibfnamefont {L.}~\bibnamefont {Sapienza}}, \bibinfo {author} {\bibfnamefont {C.-Z.}\ \bibnamefont {Zhang}}, \bibinfo {author} {\bibfnamefont {J.~V.}\ \bibnamefont {De~Miranda~Cardoso}}, \bibinfo {author} {\bibfnamefont {V.}~\bibnamefont {Verma}}, \bibinfo {author} {\bibfnamefont {R.}~\bibnamefont {Mirin}}, \bibinfo {author} {\bibfnamefont {S.~W.}\ \bibnamefont {Nam}}, \bibinfo {author} {\bibfnamefont {L.}~\bibnamefont {Liu}},\ and\ \bibinfo {author} {\bibfnamefont {K.}~\bibnamefont {Srinivasan}},\ }\bibfield  {journal} {\bibinfo  {journal} {Nature Communications}\ }\textbf {\bibinfo {volume} {8}},\ \href {https://doi.org/10.1038/s41467-017-00987-6} {10.1038/s41467-017-00987-6} (\bibinfo {year} {2017})\BibitemShut {NoStop}%
\bibitem [{\citenamefont {Chanana}\ \emph {et~al.}(2022)\citenamefont {Chanana}, \citenamefont {Larocque}, \citenamefont {Moreira}, \citenamefont {Carolan}, \citenamefont {Guha}, \citenamefont {Melo}, \citenamefont {Anant}, \citenamefont {Song}, \citenamefont {Englund}, \citenamefont {Blumenthal}, \citenamefont {Srinivasan},\ and\ \citenamefont {Davanco}}]{Chanana2022}%
  \BibitemOpen
  \bibfield  {author} {\bibinfo {author} {\bibfnamefont {A.}~\bibnamefont {Chanana}}, \bibinfo {author} {\bibfnamefont {H.}~\bibnamefont {Larocque}}, \bibinfo {author} {\bibfnamefont {R.}~\bibnamefont {Moreira}}, \bibinfo {author} {\bibfnamefont {J.}~\bibnamefont {Carolan}}, \bibinfo {author} {\bibfnamefont {B.}~\bibnamefont {Guha}}, \bibinfo {author} {\bibfnamefont {E.~G.}\ \bibnamefont {Melo}}, \bibinfo {author} {\bibfnamefont {V.}~\bibnamefont {Anant}}, \bibinfo {author} {\bibfnamefont {J.}~\bibnamefont {Song}}, \bibinfo {author} {\bibfnamefont {D.}~\bibnamefont {Englund}}, \bibinfo {author} {\bibfnamefont {D.~J.}\ \bibnamefont {Blumenthal}}, \bibinfo {author} {\bibfnamefont {K.}~\bibnamefont {Srinivasan}},\ and\ \bibinfo {author} {\bibfnamefont {M.}~\bibnamefont {Davanco}},\ }\bibfield  {journal} {\bibinfo  {journal} {Nature Communications}\ }\textbf {\bibinfo {volume} {13}},\ \href {https://doi.org/10.1038/s41467-022-35332-z} {10.1038/s41467-022-35332-z} (\bibinfo {year} {2022})\BibitemShut {NoStop}%
\bibitem [{\citenamefont {Zhu}\ \emph {et~al.}(2022)\citenamefont {Zhu}, \citenamefont {Wei}, \citenamefont {Yi}, \citenamefont {Jin}, \citenamefont {Shen}, \citenamefont {Wang}, \citenamefont {Zhou}, \citenamefont {Wang}, \citenamefont {Ou}, \citenamefont {Song}, \citenamefont {Wang}, \citenamefont {Zhang}, \citenamefont {Ou},\ and\ \citenamefont {Zhang}}]{Zhu2022}%
  \BibitemOpen
  \bibfield  {author} {\bibinfo {author} {\bibfnamefont {Y.}~\bibnamefont {Zhu}}, \bibinfo {author} {\bibfnamefont {W.}~\bibnamefont {Wei}}, \bibinfo {author} {\bibfnamefont {A.}~\bibnamefont {Yi}}, \bibinfo {author} {\bibfnamefont {T.}~\bibnamefont {Jin}}, \bibinfo {author} {\bibfnamefont {C.}~\bibnamefont {Shen}}, \bibinfo {author} {\bibfnamefont {X.}~\bibnamefont {Wang}}, \bibinfo {author} {\bibfnamefont {L.}~\bibnamefont {Zhou}}, \bibinfo {author} {\bibfnamefont {C.}~\bibnamefont {Wang}}, \bibinfo {author} {\bibfnamefont {W.}~\bibnamefont {Ou}}, \bibinfo {author} {\bibfnamefont {S.}~\bibnamefont {Song}}, \bibinfo {author} {\bibfnamefont {T.}~\bibnamefont {Wang}}, \bibinfo {author} {\bibfnamefont {J.}~\bibnamefont {Zhang}}, \bibinfo {author} {\bibfnamefont {X.}~\bibnamefont {Ou}},\ and\ \bibinfo {author} {\bibfnamefont {J.}~\bibnamefont {Zhang}},\ }\bibfield  {journal} {\bibinfo  {journal} {Laser \& Photonics Reviews}\ }\textbf {\bibinfo {volume} {16}},\ \href {https://doi.org/10.1002/lpor.202200172}
  {10.1002/lpor.202200172} (\bibinfo {year} {2022})\BibitemShut {NoStop}%
\bibitem [{\citenamefont {Zhu}\ \emph {et~al.}(2024)\citenamefont {Zhu}, \citenamefont {Liu}, \citenamefont {Yi}, \citenamefont {Wang}, \citenamefont {Qin}, \citenamefont {Zhao}, \citenamefont {Zhao}, \citenamefont {Chen}, \citenamefont {Zhang}, \citenamefont {Song}, \citenamefont {Huo}, \citenamefont {Ou},\ and\ \citenamefont {Zhang}}]{zhu2024}%
  \BibitemOpen
  \bibfield  {author} {\bibinfo {author} {\bibfnamefont {Y.}~\bibnamefont {Zhu}}, \bibinfo {author} {\bibfnamefont {R.}~\bibnamefont {Liu}}, \bibinfo {author} {\bibfnamefont {A.}~\bibnamefont {Yi}}, \bibinfo {author} {\bibfnamefont {X.}~\bibnamefont {Wang}}, \bibinfo {author} {\bibfnamefont {Y.}~\bibnamefont {Qin}}, \bibinfo {author} {\bibfnamefont {Z.}~\bibnamefont {Zhao}}, \bibinfo {author} {\bibfnamefont {J.}~\bibnamefont {Zhao}}, \bibinfo {author} {\bibfnamefont {B.}~\bibnamefont {Chen}}, \bibinfo {author} {\bibfnamefont {X.}~\bibnamefont {Zhang}}, \bibinfo {author} {\bibfnamefont {S.}~\bibnamefont {Song}}, \bibinfo {author} {\bibfnamefont {Y.}~\bibnamefont {Huo}}, \bibinfo {author} {\bibfnamefont {X.}~\bibnamefont {Ou}},\ and\ \bibinfo {author} {\bibfnamefont {J.}~\bibnamefont {Zhang}},\ }\bibfield  {journal} {\bibinfo  {journal} {arXiv}\ }\href {https://doi.org/10.48550/ARXIV.2411.06677} {10.48550/ARXIV.2411.06677} (\bibinfo {year} {2024})\BibitemShut {NoStop}%
\bibitem [{\citenamefont {Aghaeimeibodi}\ \emph {et~al.}(2018)\citenamefont {Aghaeimeibodi}, \citenamefont {Desiatov}, \citenamefont {Kim}, \citenamefont {Lee}, \citenamefont {Buyukkaya}, \citenamefont {Karasahin}, \citenamefont {Richardson}, \citenamefont {Leavitt}, \citenamefont {Lončar},\ and\ \citenamefont {Waks}}]{Aghaeimeibodi2018}%
  \BibitemOpen
  \bibfield  {author} {\bibinfo {author} {\bibfnamefont {S.}~\bibnamefont {Aghaeimeibodi}}, \bibinfo {author} {\bibfnamefont {B.}~\bibnamefont {Desiatov}}, \bibinfo {author} {\bibfnamefont {J.-H.}\ \bibnamefont {Kim}}, \bibinfo {author} {\bibfnamefont {C.-M.}\ \bibnamefont {Lee}}, \bibinfo {author} {\bibfnamefont {M.~A.}\ \bibnamefont {Buyukkaya}}, \bibinfo {author} {\bibfnamefont {A.}~\bibnamefont {Karasahin}}, \bibinfo {author} {\bibfnamefont {C.~J.~K.}\ \bibnamefont {Richardson}}, \bibinfo {author} {\bibfnamefont {R.~P.}\ \bibnamefont {Leavitt}}, \bibinfo {author} {\bibfnamefont {M.}~\bibnamefont {Lončar}},\ and\ \bibinfo {author} {\bibfnamefont {E.}~\bibnamefont {Waks}},\ }\bibfield  {journal} {\bibinfo  {journal} {Applied Physics Letters}\ }\textbf {\bibinfo {volume} {113}},\ \href {https://doi.org/10.1063/1.5054865} {10.1063/1.5054865} (\bibinfo {year} {2018})\BibitemShut {NoStop}%
\bibitem [{\citenamefont {Grim}\ \emph {et~al.}(2019)\citenamefont {Grim}, \citenamefont {Bracker}, \citenamefont {Zalalutdinov}, \citenamefont {Carter}, \citenamefont {Kozen}, \citenamefont {Kim}, \citenamefont {Kim}, \citenamefont {Mlack}, \citenamefont {Yakes}, \citenamefont {Lee},\ and\ \citenamefont {Gammon}}]{Grim2019}%
  \BibitemOpen
  \bibfield  {author} {\bibinfo {author} {\bibfnamefont {J.~Q.}\ \bibnamefont {Grim}}, \bibinfo {author} {\bibfnamefont {A.~S.}\ \bibnamefont {Bracker}}, \bibinfo {author} {\bibfnamefont {M.}~\bibnamefont {Zalalutdinov}}, \bibinfo {author} {\bibfnamefont {S.~G.}\ \bibnamefont {Carter}}, \bibinfo {author} {\bibfnamefont {A.~C.}\ \bibnamefont {Kozen}}, \bibinfo {author} {\bibfnamefont {M.}~\bibnamefont {Kim}}, \bibinfo {author} {\bibfnamefont {C.~S.}\ \bibnamefont {Kim}}, \bibinfo {author} {\bibfnamefont {J.~T.}\ \bibnamefont {Mlack}}, \bibinfo {author} {\bibfnamefont {M.}~\bibnamefont {Yakes}}, \bibinfo {author} {\bibfnamefont {B.}~\bibnamefont {Lee}},\ and\ \bibinfo {author} {\bibfnamefont {D.}~\bibnamefont {Gammon}},\ }\href {https://doi.org/10.1038/s41563-019-0418-0} {\bibfield  {journal} {\bibinfo  {journal} {Nature Materials}\ }\textbf {\bibinfo {volume} {18}},\ \bibinfo {pages} {963–969} (\bibinfo {year} {2019})}\BibitemShut {NoStop}%
\bibitem [{\citenamefont {Patel}\ \emph {et~al.}(2010)\citenamefont {Patel}, \citenamefont {Bennett}, \citenamefont {Farrer}, \citenamefont {Nicoll}, \citenamefont {Ritchie},\ and\ \citenamefont {Shields}}]{Patel2010}%
  \BibitemOpen
  \bibfield  {author} {\bibinfo {author} {\bibfnamefont {R.~B.}\ \bibnamefont {Patel}}, \bibinfo {author} {\bibfnamefont {A.~J.}\ \bibnamefont {Bennett}}, \bibinfo {author} {\bibfnamefont {I.}~\bibnamefont {Farrer}}, \bibinfo {author} {\bibfnamefont {C.~A.}\ \bibnamefont {Nicoll}}, \bibinfo {author} {\bibfnamefont {D.~A.}\ \bibnamefont {Ritchie}},\ and\ \bibinfo {author} {\bibfnamefont {A.~J.}\ \bibnamefont {Shields}},\ }\href {https://doi.org/10.1038/nphoton.2010.161} {\bibfield  {journal} {\bibinfo  {journal} {Nature Photonics}\ }\textbf {\bibinfo {volume} {4}},\ \bibinfo {pages} {632–635} (\bibinfo {year} {2010})}\BibitemShut {NoStop}%
\bibitem [{\citenamefont {Jin}\ \emph {et~al.}(2022)\citenamefont {Jin}, \citenamefont {Li}, \citenamefont {Liu}, \citenamefont {Ou}, \citenamefont {Zhu}, \citenamefont {Wang}, \citenamefont {Liu}, \citenamefont {Huo}, \citenamefont {Ou},\ and\ \citenamefont {Zhang}}]{Jin2022}%
  \BibitemOpen
  \bibfield  {author} {\bibinfo {author} {\bibfnamefont {T.}~\bibnamefont {Jin}}, \bibinfo {author} {\bibfnamefont {X.}~\bibnamefont {Li}}, \bibinfo {author} {\bibfnamefont {R.}~\bibnamefont {Liu}}, \bibinfo {author} {\bibfnamefont {W.}~\bibnamefont {Ou}}, \bibinfo {author} {\bibfnamefont {Y.}~\bibnamefont {Zhu}}, \bibinfo {author} {\bibfnamefont {X.}~\bibnamefont {Wang}}, \bibinfo {author} {\bibfnamefont {J.}~\bibnamefont {Liu}}, \bibinfo {author} {\bibfnamefont {Y.}~\bibnamefont {Huo}}, \bibinfo {author} {\bibfnamefont {X.}~\bibnamefont {Ou}},\ and\ \bibinfo {author} {\bibfnamefont {J.}~\bibnamefont {Zhang}},\ }\href {https://doi.org/10.1021/acs.nanolett.1c03226} {\bibfield  {journal} {\bibinfo  {journal} {Nano Letters}\ }\textbf {\bibinfo {volume} {22}},\ \bibinfo {pages} {586–593} (\bibinfo {year} {2022})}\BibitemShut {NoStop}%
\bibitem [{\citenamefont {Elshaari}\ \emph {et~al.}(2018)\citenamefont {Elshaari}, \citenamefont {B\"{u}y\"{u}k\"{o}zer}, \citenamefont {Zadeh}, \citenamefont {Lettner}, \citenamefont {Zhao}, \citenamefont {Sch\"{o}ll}, \citenamefont {Gyger}, \citenamefont {Reimer}, \citenamefont {Dalacu}, \citenamefont {Poole}, \citenamefont {J\"{o}ns},\ and\ \citenamefont {Zwiller}}]{Elshaari2018}%
  \BibitemOpen
  \bibfield  {author} {\bibinfo {author} {\bibfnamefont {A.~W.}\ \bibnamefont {Elshaari}}, \bibinfo {author} {\bibfnamefont {E.}~\bibnamefont {B\"{u}y\"{u}k\"{o}zer}}, \bibinfo {author} {\bibfnamefont {I.~E.}\ \bibnamefont {Zadeh}}, \bibinfo {author} {\bibfnamefont {T.}~\bibnamefont {Lettner}}, \bibinfo {author} {\bibfnamefont {P.}~\bibnamefont {Zhao}}, \bibinfo {author} {\bibfnamefont {E.}~\bibnamefont {Sch\"{o}ll}}, \bibinfo {author} {\bibfnamefont {S.}~\bibnamefont {Gyger}}, \bibinfo {author} {\bibfnamefont {M.~E.}\ \bibnamefont {Reimer}}, \bibinfo {author} {\bibfnamefont {D.}~\bibnamefont {Dalacu}}, \bibinfo {author} {\bibfnamefont {P.~J.}\ \bibnamefont {Poole}}, \bibinfo {author} {\bibfnamefont {K.~D.}\ \bibnamefont {J\"{o}ns}},\ and\ \bibinfo {author} {\bibfnamefont {V.}~\bibnamefont {Zwiller}},\ }\href {https://doi.org/10.1021/acs.nanolett.8b03937} {\bibfield  {journal} {\bibinfo  {journal} {Nano Letters}\ }\textbf {\bibinfo {volume} {18}},\ \bibinfo {pages} {7969–7976} (\bibinfo {year}
  {2018})}\BibitemShut {NoStop}%
\bibitem [{\citenamefont {Tao}\ \emph {et~al.}(2020)\citenamefont {Tao}, \citenamefont {Wei}, \citenamefont {Li}, \citenamefont {Ou}, \citenamefont {Wang}, \citenamefont {Wang}, \citenamefont {Zhang}, \citenamefont {Zhang}, \citenamefont {Gan},\ and\ \citenamefont {Ou}}]{Tao2020}%
  \BibitemOpen
  \bibfield  {author} {\bibinfo {author} {\bibfnamefont {L.}~\bibnamefont {Tao}}, \bibinfo {author} {\bibfnamefont {W.}~\bibnamefont {Wei}}, \bibinfo {author} {\bibfnamefont {Y.}~\bibnamefont {Li}}, \bibinfo {author} {\bibfnamefont {W.}~\bibnamefont {Ou}}, \bibinfo {author} {\bibfnamefont {T.}~\bibnamefont {Wang}}, \bibinfo {author} {\bibfnamefont {C.}~\bibnamefont {Wang}}, \bibinfo {author} {\bibfnamefont {J.}~\bibnamefont {Zhang}}, \bibinfo {author} {\bibfnamefont {J.}~\bibnamefont {Zhang}}, \bibinfo {author} {\bibfnamefont {F.}~\bibnamefont {Gan}},\ and\ \bibinfo {author} {\bibfnamefont {X.}~\bibnamefont {Ou}},\ }\href {https://doi.org/10.1021/acsphotonics.0c00748} {\bibfield  {journal} {\bibinfo  {journal} {ACS Photonics}\ }\textbf {\bibinfo {volume} {7}},\ \bibinfo {pages} {2723–2730} (\bibinfo {year} {2020})}\BibitemShut {NoStop}%
\bibitem [{\citenamefont {Kim}\ \emph {et~al.}(2018)\citenamefont {Kim}, \citenamefont {Aghaeimeibodi}, \citenamefont {Richardson}, \citenamefont {Leavitt},\ and\ \citenamefont {Waks}}]{Kim2018}%
  \BibitemOpen
  \bibfield  {author} {\bibinfo {author} {\bibfnamefont {J.-H.}\ \bibnamefont {Kim}}, \bibinfo {author} {\bibfnamefont {S.}~\bibnamefont {Aghaeimeibodi}}, \bibinfo {author} {\bibfnamefont {C.~J.~K.}\ \bibnamefont {Richardson}}, \bibinfo {author} {\bibfnamefont {R.~P.}\ \bibnamefont {Leavitt}},\ and\ \bibinfo {author} {\bibfnamefont {E.}~\bibnamefont {Waks}},\ }\href {https://doi.org/10.1021/acs.nanolett.8b01133} {\bibfield  {journal} {\bibinfo  {journal} {Nano Letters}\ }\textbf {\bibinfo {volume} {18}},\ \bibinfo {pages} {4734–4740} (\bibinfo {year} {2018})}\BibitemShut {NoStop}%
\bibitem [{\citenamefont {Katsumi}\ \emph {et~al.}(2018)\citenamefont {Katsumi}, \citenamefont {Ota}, \citenamefont {Kakuda}, \citenamefont {Iwamoto},\ and\ \citenamefont {Arakawa}}]{Katsumi2018}%
  \BibitemOpen
  \bibfield  {author} {\bibinfo {author} {\bibfnamefont {R.}~\bibnamefont {Katsumi}}, \bibinfo {author} {\bibfnamefont {Y.}~\bibnamefont {Ota}}, \bibinfo {author} {\bibfnamefont {M.}~\bibnamefont {Kakuda}}, \bibinfo {author} {\bibfnamefont {S.}~\bibnamefont {Iwamoto}},\ and\ \bibinfo {author} {\bibfnamefont {Y.}~\bibnamefont {Arakawa}},\ }\href {https://doi.org/10.1364/optica.5.000691} {\bibfield  {journal} {\bibinfo  {journal} {Optica}\ }\textbf {\bibinfo {volume} {5}},\ \bibinfo {pages} {691} (\bibinfo {year} {2018})}\BibitemShut {NoStop}%
\bibitem [{\citenamefont {Osada}\ \emph {et~al.}(2019)\citenamefont {Osada}, \citenamefont {Ota}, \citenamefont {Katsumi}, \citenamefont {Kakuda}, \citenamefont {Iwamoto},\ and\ \citenamefont {Arakawa}}]{Osada2019}%
  \BibitemOpen
  \bibfield  {author} {\bibinfo {author} {\bibfnamefont {A.}~\bibnamefont {Osada}}, \bibinfo {author} {\bibfnamefont {Y.}~\bibnamefont {Ota}}, \bibinfo {author} {\bibfnamefont {R.}~\bibnamefont {Katsumi}}, \bibinfo {author} {\bibfnamefont {M.}~\bibnamefont {Kakuda}}, \bibinfo {author} {\bibfnamefont {S.}~\bibnamefont {Iwamoto}},\ and\ \bibinfo {author} {\bibfnamefont {Y.}~\bibnamefont {Arakawa}},\ }\bibfield  {journal} {\bibinfo  {journal} {Physical Review Applied}\ }\textbf {\bibinfo {volume} {11}},\ \href {https://doi.org/10.1103/physrevapplied.11.024071} {10.1103/physrevapplied.11.024071} (\bibinfo {year} {2019})\BibitemShut {NoStop}%
\bibitem [{\citenamefont {Katsumi}\ \emph {et~al.}(2019)\citenamefont {Katsumi}, \citenamefont {Ota}, \citenamefont {Osada}, \citenamefont {Yamaguchi}, \citenamefont {Tajiri}, \citenamefont {Kakuda}, \citenamefont {Iwamoto}, \citenamefont {Akiyama},\ and\ \citenamefont {Arakawa}}]{Katsumi2019}%
  \BibitemOpen
  \bibfield  {author} {\bibinfo {author} {\bibfnamefont {R.}~\bibnamefont {Katsumi}}, \bibinfo {author} {\bibfnamefont {Y.}~\bibnamefont {Ota}}, \bibinfo {author} {\bibfnamefont {A.}~\bibnamefont {Osada}}, \bibinfo {author} {\bibfnamefont {T.}~\bibnamefont {Yamaguchi}}, \bibinfo {author} {\bibfnamefont {T.}~\bibnamefont {Tajiri}}, \bibinfo {author} {\bibfnamefont {M.}~\bibnamefont {Kakuda}}, \bibinfo {author} {\bibfnamefont {S.}~\bibnamefont {Iwamoto}}, \bibinfo {author} {\bibfnamefont {H.}~\bibnamefont {Akiyama}},\ and\ \bibinfo {author} {\bibfnamefont {Y.}~\bibnamefont {Arakawa}},\ }\bibfield  {journal} {\bibinfo  {journal} {APL Photonics}\ }\textbf {\bibinfo {volume} {4}},\ \href {https://doi.org/10.1063/1.5087263} {10.1063/1.5087263} (\bibinfo {year} {2019})\BibitemShut {NoStop}%
\bibitem [{\citenamefont {Katsumi}\ \emph {et~al.}(2020)\citenamefont {Katsumi}, \citenamefont {Ota}, \citenamefont {Osada}, \citenamefont {Tajiri}, \citenamefont {Yamaguchi}, \citenamefont {Kakuda}, \citenamefont {Iwamoto}, \citenamefont {Akiyama},\ and\ \citenamefont {Arakawa}}]{Katsumi2020}%
  \BibitemOpen
  \bibfield  {author} {\bibinfo {author} {\bibfnamefont {R.}~\bibnamefont {Katsumi}}, \bibinfo {author} {\bibfnamefont {Y.}~\bibnamefont {Ota}}, \bibinfo {author} {\bibfnamefont {A.}~\bibnamefont {Osada}}, \bibinfo {author} {\bibfnamefont {T.}~\bibnamefont {Tajiri}}, \bibinfo {author} {\bibfnamefont {T.}~\bibnamefont {Yamaguchi}}, \bibinfo {author} {\bibfnamefont {M.}~\bibnamefont {Kakuda}}, \bibinfo {author} {\bibfnamefont {S.}~\bibnamefont {Iwamoto}}, \bibinfo {author} {\bibfnamefont {H.}~\bibnamefont {Akiyama}},\ and\ \bibinfo {author} {\bibfnamefont {Y.}~\bibnamefont {Arakawa}},\ }\bibfield  {journal} {\bibinfo  {journal} {Applied Physics Letters}\ }\textbf {\bibinfo {volume} {116}},\ \href {https://doi.org/10.1063/1.5129325} {10.1063/1.5129325} (\bibinfo {year} {2020})\BibitemShut {NoStop}%
\bibitem [{\citenamefont {Papon}\ \emph {et~al.}(2023)\citenamefont {Papon}, \citenamefont {Wang}, \citenamefont {Uppu}, \citenamefont {Scholz}, \citenamefont {Wieck}, \citenamefont {Ludwig}, \citenamefont {Lodahl},\ and\ \citenamefont {Midolo}}]{Papon2023}%
  \BibitemOpen
  \bibfield  {author} {\bibinfo {author} {\bibfnamefont {C.}~\bibnamefont {Papon}}, \bibinfo {author} {\bibfnamefont {Y.}~\bibnamefont {Wang}}, \bibinfo {author} {\bibfnamefont {R.}~\bibnamefont {Uppu}}, \bibinfo {author} {\bibfnamefont {S.}~\bibnamefont {Scholz}}, \bibinfo {author} {\bibfnamefont {A.}~\bibnamefont {Wieck}}, \bibinfo {author} {\bibfnamefont {A.}~\bibnamefont {Ludwig}}, \bibinfo {author} {\bibfnamefont {P.}~\bibnamefont {Lodahl}},\ and\ \bibinfo {author} {\bibfnamefont {L.}~\bibnamefont {Midolo}},\ }\bibfield  {journal} {\bibinfo  {journal} {Physical Review Applied}\ }\textbf {\bibinfo {volume} {19}},\ \href {https://doi.org/10.1103/physrevapplied.19.l061003} {10.1103/physrevapplied.19.l061003} (\bibinfo {year} {2023})\BibitemShut {NoStop}%
\bibitem [{\citenamefont {Wan}\ \emph {et~al.}(2020)\citenamefont {Wan}, \citenamefont {Lu}, \citenamefont {Chen}, \citenamefont {Walsh}, \citenamefont {Trusheim}, \citenamefont {De~Santis}, \citenamefont {Bersin}, \citenamefont {Harris}, \citenamefont {Mouradian}, \citenamefont {Christen}, \citenamefont {Bielejec},\ and\ \citenamefont {Englund}}]{Wan2020}%
  \BibitemOpen
  \bibfield  {author} {\bibinfo {author} {\bibfnamefont {N.~H.}\ \bibnamefont {Wan}}, \bibinfo {author} {\bibfnamefont {T.-J.}\ \bibnamefont {Lu}}, \bibinfo {author} {\bibfnamefont {K.~C.}\ \bibnamefont {Chen}}, \bibinfo {author} {\bibfnamefont {M.~P.}\ \bibnamefont {Walsh}}, \bibinfo {author} {\bibfnamefont {M.~E.}\ \bibnamefont {Trusheim}}, \bibinfo {author} {\bibfnamefont {L.}~\bibnamefont {De~Santis}}, \bibinfo {author} {\bibfnamefont {E.~A.}\ \bibnamefont {Bersin}}, \bibinfo {author} {\bibfnamefont {I.~B.}\ \bibnamefont {Harris}}, \bibinfo {author} {\bibfnamefont {S.~L.}\ \bibnamefont {Mouradian}}, \bibinfo {author} {\bibfnamefont {I.~R.}\ \bibnamefont {Christen}}, \bibinfo {author} {\bibfnamefont {E.~S.}\ \bibnamefont {Bielejec}},\ and\ \bibinfo {author} {\bibfnamefont {D.}~\bibnamefont {Englund}},\ }\href {https://doi.org/10.1038/s41586-020-2441-3} {\bibfield  {journal} {\bibinfo  {journal} {Nature}\ }\textbf {\bibinfo {volume} {583}},\ \bibinfo {pages} {226–231} (\bibinfo {year} {2020})}\BibitemShut
  {NoStop}%
\bibitem [{\citenamefont {Li}\ \emph {et~al.}(2024)\citenamefont {Li}, \citenamefont {Santis}, \citenamefont {Harris}, \citenamefont {Chen}, \citenamefont {Gao}, \citenamefont {Christen}, \citenamefont {Choi}, \citenamefont {Trusheim}, \citenamefont {Song}, \citenamefont {Errando-Herranz}, \citenamefont {Du}, \citenamefont {Hu}, \citenamefont {Clark}, \citenamefont {Ibrahim}, \citenamefont {Gilbert}, \citenamefont {Han},\ and\ \citenamefont {Englund}}]{Li2024}%
  \BibitemOpen
  \bibfield  {author} {\bibinfo {author} {\bibfnamefont {L.}~\bibnamefont {Li}}, \bibinfo {author} {\bibfnamefont {L.~D.}\ \bibnamefont {Santis}}, \bibinfo {author} {\bibfnamefont {I.~B.~W.}\ \bibnamefont {Harris}}, \bibinfo {author} {\bibfnamefont {K.~C.}\ \bibnamefont {Chen}}, \bibinfo {author} {\bibfnamefont {Y.}~\bibnamefont {Gao}}, \bibinfo {author} {\bibfnamefont {I.}~\bibnamefont {Christen}}, \bibinfo {author} {\bibfnamefont {H.}~\bibnamefont {Choi}}, \bibinfo {author} {\bibfnamefont {M.}~\bibnamefont {Trusheim}}, \bibinfo {author} {\bibfnamefont {Y.}~\bibnamefont {Song}}, \bibinfo {author} {\bibfnamefont {C.}~\bibnamefont {Errando-Herranz}}, \bibinfo {author} {\bibfnamefont {J.}~\bibnamefont {Du}}, \bibinfo {author} {\bibfnamefont {Y.}~\bibnamefont {Hu}}, \bibinfo {author} {\bibfnamefont {G.}~\bibnamefont {Clark}}, \bibinfo {author} {\bibfnamefont {M.~I.}\ \bibnamefont {Ibrahim}}, \bibinfo {author} {\bibfnamefont {G.}~\bibnamefont {Gilbert}}, \bibinfo {author} {\bibfnamefont {R.}~\bibnamefont {Han}},\
  and\ \bibinfo {author} {\bibfnamefont {D.}~\bibnamefont {Englund}},\ }\href {https://doi.org/10.1038/s41586-024-07371-7} {\bibfield  {journal} {\bibinfo  {journal} {Nature}\ }\textbf {\bibinfo {volume} {630}},\ \bibinfo {pages} {70–76} (\bibinfo {year} {2024})}\BibitemShut {NoStop}%
\bibitem [{\citenamefont {Bukhari}\ \emph {et~al.}(2014)\citenamefont {Bukhari}, \citenamefont {Islam}, \citenamefont {Haziot},\ and\ \citenamefont {Beamish}}]{Bukhari2014}%
  \BibitemOpen
  \bibfield  {author} {\bibinfo {author} {\bibfnamefont {S.}~\bibnamefont {Bukhari}}, \bibinfo {author} {\bibfnamefont {M.}~\bibnamefont {Islam}}, \bibinfo {author} {\bibfnamefont {A.}~\bibnamefont {Haziot}},\ and\ \bibinfo {author} {\bibfnamefont {J.}~\bibnamefont {Beamish}},\ }\href {https://doi.org/10.1088/1742-6596/568/3/032004} {\bibfield  {journal} {\bibinfo  {journal} {Journal of Physics: Conference Series}\ }\textbf {\bibinfo {volume} {568}},\ \bibinfo {pages} {032004} (\bibinfo {year} {2014})}\BibitemShut {NoStop}%
\bibitem [{\citenamefont {Islam}\ and\ \citenamefont {Beamish}(2019)}]{islam2019}%
  \BibitemOpen
  \bibfield  {author} {\bibinfo {author} {\bibfnamefont {M.~S.}\ \bibnamefont {Islam}}\ and\ \bibinfo {author} {\bibfnamefont {J.}~\bibnamefont {Beamish}},\ }\href {https://doi.org/10.1063/1.5119351} {\bibfield  {journal} {\bibinfo  {journal} {Journal of Applied Physics}\ }\textbf {\bibinfo {volume} {126}},\ \bibinfo {pages} {204101} (\bibinfo {year} {2019})}\BibitemShut {NoStop}%
\bibitem [{\citenamefont {Newnham}(2004)}]{newnham2004}%
  \BibitemOpen
  \bibfield  {author} {\bibinfo {author} {\bibfnamefont {R.~E.}\ \bibnamefont {Newnham}},\ }\href {https://doi.org/10.1093/oso/9780198520757.003.0003} {\emph {\bibinfo {title} {Properties of {{Materials Anisotropy}}, {{Symmetry}}, {{Structure}}}}},\ edited by\ \bibinfo {editor} {\bibfnamefont {R.~E.}\ \bibnamefont {Newnham}}\ (\bibinfo  {publisher} {Oxford University Press},\ \bibinfo {year} {2004})\BibitemShut {NoStop}%
\bibitem [{\citenamefont {Tarumi}\ \emph {et~al.}(2012)\citenamefont {Tarumi}, \citenamefont {Matsuhisa},\ and\ \citenamefont {Shibutani}}]{tarumi2012}%
  \BibitemOpen
  \bibfield  {author} {\bibinfo {author} {\bibfnamefont {R.}~\bibnamefont {Tarumi}}, \bibinfo {author} {\bibfnamefont {T.}~\bibnamefont {Matsuhisa}},\ and\ \bibinfo {author} {\bibfnamefont {Y.}~\bibnamefont {Shibutani}},\ }\href {https://doi.org/10.1143/JJAP.51.07GA02} {\bibfield  {journal} {\bibinfo  {journal} {Japanese Journal of Applied Physics}\ }\textbf {\bibinfo {volume} {51}},\ \bibinfo {pages} {07GA02} (\bibinfo {year} {2012})}\BibitemShut {NoStop}%
\bibitem [{\citenamefont {Sun}\ \emph {et~al.}(2010)\citenamefont {Sun}, \citenamefont {Thompson},\ and\ \citenamefont {Nishida}}]{Sun2010}%
  \BibitemOpen
  \bibfield  {author} {\bibinfo {author} {\bibfnamefont {Y.}~\bibnamefont {Sun}}, \bibinfo {author} {\bibfnamefont {S.~E.}\ \bibnamefont {Thompson}},\ and\ \bibinfo {author} {\bibfnamefont {T.}~\bibnamefont {Nishida}},\ }\href {https://doi.org/10.1007/978-1-4419-0552-9} {\emph {\bibinfo {title} {Strain Effect in Semiconductors}}}\ (\bibinfo  {publisher} {Springer US},\ \bibinfo {year} {2010})\BibitemShut {NoStop}%
\bibitem [{\citenamefont {Kuhlmann}\ \emph {et~al.}(2015)\citenamefont {Kuhlmann}, \citenamefont {Prechtel}, \citenamefont {Houel}, \citenamefont {Ludwig}, \citenamefont {Reuter}, \citenamefont {Wieck},\ and\ \citenamefont {Warburton}}]{Kuhlmann2015}%
  \BibitemOpen
  \bibfield  {author} {\bibinfo {author} {\bibfnamefont {A.~V.}\ \bibnamefont {Kuhlmann}}, \bibinfo {author} {\bibfnamefont {J.~H.}\ \bibnamefont {Prechtel}}, \bibinfo {author} {\bibfnamefont {J.}~\bibnamefont {Houel}}, \bibinfo {author} {\bibfnamefont {A.}~\bibnamefont {Ludwig}}, \bibinfo {author} {\bibfnamefont {D.}~\bibnamefont {Reuter}}, \bibinfo {author} {\bibfnamefont {A.~D.}\ \bibnamefont {Wieck}},\ and\ \bibinfo {author} {\bibfnamefont {R.~J.}\ \bibnamefont {Warburton}},\ }\bibfield  {journal} {\bibinfo  {journal} {Nature Communications}\ }\textbf {\bibinfo {volume} {6}},\ \href {https://doi.org/10.1038/ncomms9204} {10.1038/ncomms9204} (\bibinfo {year} {2015})\BibitemShut {NoStop}%
\bibitem [{\citenamefont {Thyrrestrup}\ \emph {et~al.}(2018)\citenamefont {Thyrrestrup}, \citenamefont {Kiršanskė}, \citenamefont {Le~Jeannic}, \citenamefont {Pregnolato}, \citenamefont {Zhai}, \citenamefont {Raahauge}, \citenamefont {Midolo}, \citenamefont {Rotenberg}, \citenamefont {Javadi}, \citenamefont {Schott}, \citenamefont {Wieck}, \citenamefont {Ludwig}, \citenamefont {L\"{o}bl}, \citenamefont {S\"{o}llner}, \citenamefont {Warburton},\ and\ \citenamefont {Lodahl}}]{Thyrrestrup2018}%
  \BibitemOpen
  \bibfield  {author} {\bibinfo {author} {\bibfnamefont {H.}~\bibnamefont {Thyrrestrup}}, \bibinfo {author} {\bibfnamefont {G.}~\bibnamefont {Kiršanskė}}, \bibinfo {author} {\bibfnamefont {H.}~\bibnamefont {Le~Jeannic}}, \bibinfo {author} {\bibfnamefont {T.}~\bibnamefont {Pregnolato}}, \bibinfo {author} {\bibfnamefont {L.}~\bibnamefont {Zhai}}, \bibinfo {author} {\bibfnamefont {L.}~\bibnamefont {Raahauge}}, \bibinfo {author} {\bibfnamefont {L.}~\bibnamefont {Midolo}}, \bibinfo {author} {\bibfnamefont {N.}~\bibnamefont {Rotenberg}}, \bibinfo {author} {\bibfnamefont {A.}~\bibnamefont {Javadi}}, \bibinfo {author} {\bibfnamefont {R.}~\bibnamefont {Schott}}, \bibinfo {author} {\bibfnamefont {A.~D.}\ \bibnamefont {Wieck}}, \bibinfo {author} {\bibfnamefont {A.}~\bibnamefont {Ludwig}}, \bibinfo {author} {\bibfnamefont {M.~C.}\ \bibnamefont {L\"{o}bl}}, \bibinfo {author} {\bibfnamefont {I.}~\bibnamefont {S\"{o}llner}}, \bibinfo {author} {\bibfnamefont {R.~J.}\ \bibnamefont {Warburton}},\ and\ \bibinfo {author}
  {\bibfnamefont {P.}~\bibnamefont {Lodahl}},\ }\href {https://doi.org/10.1021/acs.nanolett.7b05016} {\bibfield  {journal} {\bibinfo  {journal} {Nano Letters}\ }\textbf {\bibinfo {volume} {18}},\ \bibinfo {pages} {1801–1806} (\bibinfo {year} {2018})}\BibitemShut {NoStop}%
\bibitem [{\citenamefont {Zhang}\ \emph {et~al.}(2013)\citenamefont {Zhang}, \citenamefont {Ding}, \citenamefont {Zallo}, \citenamefont {Trotta}, \citenamefont {H\"{o}fer}, \citenamefont {Han}, \citenamefont {Kumar}, \citenamefont {Huo}, \citenamefont {Rastelli},\ and\ \citenamefont {Schmidt}}]{Zhang2013}%
  \BibitemOpen
  \bibfield  {author} {\bibinfo {author} {\bibfnamefont {J.}~\bibnamefont {Zhang}}, \bibinfo {author} {\bibfnamefont {F.}~\bibnamefont {Ding}}, \bibinfo {author} {\bibfnamefont {E.}~\bibnamefont {Zallo}}, \bibinfo {author} {\bibfnamefont {R.}~\bibnamefont {Trotta}}, \bibinfo {author} {\bibfnamefont {B.}~\bibnamefont {H\"{o}fer}}, \bibinfo {author} {\bibfnamefont {L.}~\bibnamefont {Han}}, \bibinfo {author} {\bibfnamefont {S.}~\bibnamefont {Kumar}}, \bibinfo {author} {\bibfnamefont {Y.}~\bibnamefont {Huo}}, \bibinfo {author} {\bibfnamefont {A.}~\bibnamefont {Rastelli}},\ and\ \bibinfo {author} {\bibfnamefont {O.~G.}\ \bibnamefont {Schmidt}},\ }\href {https://doi.org/10.1021/nl402307q} {\bibfield  {journal} {\bibinfo  {journal} {Nano Letters}\ }\textbf {\bibinfo {volume} {13}},\ \bibinfo {pages} {5808–5813} (\bibinfo {year} {2013})}\BibitemShut {NoStop}%
\bibitem [{\citenamefont {Trotta}\ \emph {et~al.}(2012)\citenamefont {Trotta}, \citenamefont {Zallo}, \citenamefont {Ortix}, \citenamefont {Atkinson}, \citenamefont {Plumhof}, \citenamefont {van~den Brink}, \citenamefont {Rastelli},\ and\ \citenamefont {Schmidt}}]{Trotta2012}%
  \BibitemOpen
  \bibfield  {author} {\bibinfo {author} {\bibfnamefont {R.}~\bibnamefont {Trotta}}, \bibinfo {author} {\bibfnamefont {E.}~\bibnamefont {Zallo}}, \bibinfo {author} {\bibfnamefont {C.}~\bibnamefont {Ortix}}, \bibinfo {author} {\bibfnamefont {P.}~\bibnamefont {Atkinson}}, \bibinfo {author} {\bibfnamefont {J.~D.}\ \bibnamefont {Plumhof}}, \bibinfo {author} {\bibfnamefont {J.}~\bibnamefont {van~den Brink}}, \bibinfo {author} {\bibfnamefont {A.}~\bibnamefont {Rastelli}},\ and\ \bibinfo {author} {\bibfnamefont {O.~G.}\ \bibnamefont {Schmidt}},\ }\bibfield  {journal} {\bibinfo  {journal} {Physical Review Letters}\ }\textbf {\bibinfo {volume} {109}},\ \href {https://doi.org/10.1103/physrevlett.109.147401} {10.1103/physrevlett.109.147401} (\bibinfo {year} {2012})\BibitemShut {NoStop}%
\bibitem [{\citenamefont {Kim}\ \emph {et~al.}(2016)\citenamefont {Kim}, \citenamefont {Richardson}, \citenamefont {Leavitt},\ and\ \citenamefont {Waks}}]{kim2016a}%
  \BibitemOpen
  \bibfield  {author} {\bibinfo {author} {\bibfnamefont {J.-H.}\ \bibnamefont {Kim}}, \bibinfo {author} {\bibfnamefont {C.~J.~K.}\ \bibnamefont {Richardson}}, \bibinfo {author} {\bibfnamefont {R.~P.}\ \bibnamefont {Leavitt}},\ and\ \bibinfo {author} {\bibfnamefont {E.}~\bibnamefont {Waks}},\ }\href {https://doi.org/10.1021/acs.nanolett.6b03295} {\bibfield  {journal} {\bibinfo  {journal} {Nano Letters}\ }\textbf {\bibinfo {volume} {16}},\ \bibinfo {pages} {7061} (\bibinfo {year} {2016})}\BibitemShut {NoStop}%
\bibitem [{\citenamefont {{Zhe-Yu Jeff Ou}}(2007)}]{zhe-yu2007}%
  \BibitemOpen
  \bibfield  {author} {\bibinfo {author} {\bibnamefont {{Zhe-Yu Jeff Ou}}},\ }\href {https://doi.org/10.1007/978-0-387-25554-5} {\emph {\bibinfo {title} {Multi-{{Photon Quantum Interference}}}}}\ (\bibinfo  {publisher} {Springer US},\ \bibinfo {address} {Boston, MA},\ \bibinfo {year} {2007})\BibitemShut {NoStop}%
\bibitem [{\citenamefont {Stevenson}\ \emph {et~al.}(2013)\citenamefont {Stevenson}, \citenamefont {Nilsson}, \citenamefont {Bennett}, \citenamefont {{Skiba-Szymanska}}, \citenamefont {Farrer}, \citenamefont {Ritchie},\ and\ \citenamefont {Shields}}]{stevenson2013a}%
  \BibitemOpen
  \bibfield  {author} {\bibinfo {author} {\bibfnamefont {R.~M.}\ \bibnamefont {Stevenson}}, \bibinfo {author} {\bibfnamefont {J.}~\bibnamefont {Nilsson}}, \bibinfo {author} {\bibfnamefont {A.~J.}\ \bibnamefont {Bennett}}, \bibinfo {author} {\bibfnamefont {J.}~\bibnamefont {{Skiba-Szymanska}}}, \bibinfo {author} {\bibfnamefont {I.}~\bibnamefont {Farrer}}, \bibinfo {author} {\bibfnamefont {D.~A.}\ \bibnamefont {Ritchie}},\ and\ \bibinfo {author} {\bibfnamefont {A.~J.}\ \bibnamefont {Shields}},\ }\href {https://doi.org/10.1038/ncomms3859} {\bibfield  {journal} {\bibinfo  {journal} {Nature Communications}\ }\textbf {\bibinfo {volume} {4}},\ \bibinfo {pages} {2859} (\bibinfo {year} {2013})}\BibitemShut {NoStop}%
\bibitem [{\citenamefont {Patel}\ \emph {et~al.}(2008)\citenamefont {Patel}, \citenamefont {Bennett}, \citenamefont {Cooper}, \citenamefont {Atkinson}, \citenamefont {Nicoll}, \citenamefont {Ritchie},\ and\ \citenamefont {Shields}}]{Patel2008}%
  \BibitemOpen
  \bibfield  {author} {\bibinfo {author} {\bibfnamefont {R.~B.}\ \bibnamefont {Patel}}, \bibinfo {author} {\bibfnamefont {A.~J.}\ \bibnamefont {Bennett}}, \bibinfo {author} {\bibfnamefont {K.}~\bibnamefont {Cooper}}, \bibinfo {author} {\bibfnamefont {P.}~\bibnamefont {Atkinson}}, \bibinfo {author} {\bibfnamefont {C.~A.}\ \bibnamefont {Nicoll}}, \bibinfo {author} {\bibfnamefont {D.~A.}\ \bibnamefont {Ritchie}},\ and\ \bibinfo {author} {\bibfnamefont {A.~J.}\ \bibnamefont {Shields}},\ }\bibfield  {journal} {\bibinfo  {journal} {Physical Review Letters}\ }\textbf {\bibinfo {volume} {100}},\ \href {https://doi.org/10.1103/physrevlett.100.207405} {10.1103/physrevlett.100.207405} (\bibinfo {year} {2008})\BibitemShut {NoStop}%
\bibitem [{\citenamefont {Sund}\ \emph {et~al.}(2023)\citenamefont {Sund}, \citenamefont {Lomonte}, \citenamefont {Paesani}, \citenamefont {Wang}, \citenamefont {Carolan}, \citenamefont {Bart}, \citenamefont {Wieck}, \citenamefont {Ludwig}, \citenamefont {Midolo}, \citenamefont {Pernice}, \citenamefont {Lodahl},\ and\ \citenamefont {Lenzini}}]{Sund2023}%
  \BibitemOpen
  \bibfield  {author} {\bibinfo {author} {\bibfnamefont {P.~I.}\ \bibnamefont {Sund}}, \bibinfo {author} {\bibfnamefont {E.}~\bibnamefont {Lomonte}}, \bibinfo {author} {\bibfnamefont {S.}~\bibnamefont {Paesani}}, \bibinfo {author} {\bibfnamefont {Y.}~\bibnamefont {Wang}}, \bibinfo {author} {\bibfnamefont {J.}~\bibnamefont {Carolan}}, \bibinfo {author} {\bibfnamefont {N.}~\bibnamefont {Bart}}, \bibinfo {author} {\bibfnamefont {A.~D.}\ \bibnamefont {Wieck}}, \bibinfo {author} {\bibfnamefont {A.}~\bibnamefont {Ludwig}}, \bibinfo {author} {\bibfnamefont {L.}~\bibnamefont {Midolo}}, \bibinfo {author} {\bibfnamefont {W.~H.~P.}\ \bibnamefont {Pernice}}, \bibinfo {author} {\bibfnamefont {P.}~\bibnamefont {Lodahl}},\ and\ \bibinfo {author} {\bibfnamefont {F.}~\bibnamefont {Lenzini}},\ }\bibfield  {journal} {\bibinfo  {journal} {Science Advances}\ }\textbf {\bibinfo {volume} {9}},\ \href {https://doi.org/10.1126/sciadv.adg7268} {10.1126/sciadv.adg7268} (\bibinfo {year} {2023})\BibitemShut {NoStop}%
\bibitem [{\citenamefont {Lukin}\ \emph {et~al.}(2019)\citenamefont {Lukin}, \citenamefont {Dory}, \citenamefont {Guidry}, \citenamefont {Yang}, \citenamefont {Mishra}, \citenamefont {Trivedi}, \citenamefont {Radulaski}, \citenamefont {Sun}, \citenamefont {Vercruysse}, \citenamefont {Ahn},\ and\ \citenamefont {Vučković}}]{Lukin2019}%
  \BibitemOpen
  \bibfield  {author} {\bibinfo {author} {\bibfnamefont {D.~M.}\ \bibnamefont {Lukin}}, \bibinfo {author} {\bibfnamefont {C.}~\bibnamefont {Dory}}, \bibinfo {author} {\bibfnamefont {M.~A.}\ \bibnamefont {Guidry}}, \bibinfo {author} {\bibfnamefont {K.~Y.}\ \bibnamefont {Yang}}, \bibinfo {author} {\bibfnamefont {S.~D.}\ \bibnamefont {Mishra}}, \bibinfo {author} {\bibfnamefont {R.}~\bibnamefont {Trivedi}}, \bibinfo {author} {\bibfnamefont {M.}~\bibnamefont {Radulaski}}, \bibinfo {author} {\bibfnamefont {S.}~\bibnamefont {Sun}}, \bibinfo {author} {\bibfnamefont {D.}~\bibnamefont {Vercruysse}}, \bibinfo {author} {\bibfnamefont {G.~H.}\ \bibnamefont {Ahn}},\ and\ \bibinfo {author} {\bibfnamefont {J.}~\bibnamefont {Vučković}},\ }\href {https://doi.org/10.1038/s41566-019-0556-6} {\bibfield  {journal} {\bibinfo  {journal} {Nature Photonics}\ }\textbf {\bibinfo {volume} {14}},\ \bibinfo {pages} {330–334} (\bibinfo {year} {2019})}\BibitemShut {NoStop}%
\bibitem [{\citenamefont {Greentree}\ \emph {et~al.}(2008)\citenamefont {Greentree}, \citenamefont {Fairchild}, \citenamefont {Hossain},\ and\ \citenamefont {Prawer}}]{Greentree2008}%
  \BibitemOpen
  \bibfield  {author} {\bibinfo {author} {\bibfnamefont {A.~D.}\ \bibnamefont {Greentree}}, \bibinfo {author} {\bibfnamefont {B.~A.}\ \bibnamefont {Fairchild}}, \bibinfo {author} {\bibfnamefont {F.~M.}\ \bibnamefont {Hossain}},\ and\ \bibinfo {author} {\bibfnamefont {S.}~\bibnamefont {Prawer}},\ }\href {https://doi.org/10.1016/s1369-7021(08)70176-7} {\bibfield  {journal} {\bibinfo  {journal} {Materials Today}\ }\textbf {\bibinfo {volume} {11}},\ \bibinfo {pages} {22–31} (\bibinfo {year} {2008})}\BibitemShut {NoStop}%
\bibitem [{\citenamefont {Peyskens}\ \emph {et~al.}(2019)\citenamefont {Peyskens}, \citenamefont {Chakraborty}, \citenamefont {Muneeb}, \citenamefont {Van~Thourhout},\ and\ \citenamefont {Englund}}]{Peyskens2019}%
  \BibitemOpen
  \bibfield  {author} {\bibinfo {author} {\bibfnamefont {F.}~\bibnamefont {Peyskens}}, \bibinfo {author} {\bibfnamefont {C.}~\bibnamefont {Chakraborty}}, \bibinfo {author} {\bibfnamefont {M.}~\bibnamefont {Muneeb}}, \bibinfo {author} {\bibfnamefont {D.}~\bibnamefont {Van~Thourhout}},\ and\ \bibinfo {author} {\bibfnamefont {D.}~\bibnamefont {Englund}},\ }\bibfield  {journal} {\bibinfo  {journal} {Nature Communications}\ }\textbf {\bibinfo {volume} {10}},\ \href {https://doi.org/10.1038/s41467-019-12421-0} {10.1038/s41467-019-12421-0} (\bibinfo {year} {2019})\BibitemShut {NoStop}%
\bibitem [{\citenamefont {Palacios-Berraquero}\ \emph {et~al.}(2017)\citenamefont {Palacios-Berraquero}, \citenamefont {Kara}, \citenamefont {Montblanch}, \citenamefont {Barbone}, \citenamefont {Latawiec}, \citenamefont {Yoon}, \citenamefont {Ott}, \citenamefont {Loncar}, \citenamefont {Ferrari},\ and\ \citenamefont {Atat\"{u}re}}]{PalaciosBerraquero2017}%
  \BibitemOpen
  \bibfield  {author} {\bibinfo {author} {\bibfnamefont {C.}~\bibnamefont {Palacios-Berraquero}}, \bibinfo {author} {\bibfnamefont {D.~M.}\ \bibnamefont {Kara}}, \bibinfo {author} {\bibfnamefont {A.~R.-P.}\ \bibnamefont {Montblanch}}, \bibinfo {author} {\bibfnamefont {M.}~\bibnamefont {Barbone}}, \bibinfo {author} {\bibfnamefont {P.}~\bibnamefont {Latawiec}}, \bibinfo {author} {\bibfnamefont {D.}~\bibnamefont {Yoon}}, \bibinfo {author} {\bibfnamefont {A.~K.}\ \bibnamefont {Ott}}, \bibinfo {author} {\bibfnamefont {M.}~\bibnamefont {Loncar}}, \bibinfo {author} {\bibfnamefont {A.~C.}\ \bibnamefont {Ferrari}},\ and\ \bibinfo {author} {\bibfnamefont {M.}~\bibnamefont {Atat\"{u}re}},\ }\bibfield  {journal} {\bibinfo  {journal} {Nature Communications}\ }\textbf {\bibinfo {volume} {8}},\ \href {https://doi.org/10.1038/ncomms15093} {10.1038/ncomms15093} (\bibinfo {year} {2017})\BibitemShut {NoStop}%
\bibitem [{\citenamefont {Lomonte}\ \emph {et~al.}(2021)\citenamefont {Lomonte}, \citenamefont {Wolff}, \citenamefont {Beutel}, \citenamefont {Ferrari}, \citenamefont {Schuck}, \citenamefont {Pernice},\ and\ \citenamefont {Lenzini}}]{Lomonte2021}%
  \BibitemOpen
  \bibfield  {author} {\bibinfo {author} {\bibfnamefont {E.}~\bibnamefont {Lomonte}}, \bibinfo {author} {\bibfnamefont {M.~A.}\ \bibnamefont {Wolff}}, \bibinfo {author} {\bibfnamefont {F.}~\bibnamefont {Beutel}}, \bibinfo {author} {\bibfnamefont {S.}~\bibnamefont {Ferrari}}, \bibinfo {author} {\bibfnamefont {C.}~\bibnamefont {Schuck}}, \bibinfo {author} {\bibfnamefont {W.~H.~P.}\ \bibnamefont {Pernice}},\ and\ \bibinfo {author} {\bibfnamefont {F.}~\bibnamefont {Lenzini}},\ }\bibfield  {journal} {\bibinfo  {journal} {Nature Communications}\ }\textbf {\bibinfo {volume} {12}},\ \href {https://doi.org/10.1038/s41467-021-27205-8} {10.1038/s41467-021-27205-8} (\bibinfo {year} {2021})\BibitemShut {NoStop}%
\bibitem [{\citenamefont {Varnava}\ \emph {et~al.}(2008)\citenamefont {Varnava}, \citenamefont {Browne},\ and\ \citenamefont {Rudolph}}]{Varnava2008}%
  \BibitemOpen
  \bibfield  {author} {\bibinfo {author} {\bibfnamefont {M.}~\bibnamefont {Varnava}}, \bibinfo {author} {\bibfnamefont {D.~E.}\ \bibnamefont {Browne}},\ and\ \bibinfo {author} {\bibfnamefont {T.}~\bibnamefont {Rudolph}},\ }\bibfield  {journal} {\bibinfo  {journal} {Physical Review Letters}\ }\textbf {\bibinfo {volume} {100}},\ \href {https://doi.org/10.1103/physrevlett.100.060502} {10.1103/physrevlett.100.060502} (\bibinfo {year} {2008})\BibitemShut {NoStop}%
\end{thebibliography}%
	
\begin{acknowledgements}
The authors would like to acknowledge the financial support from the National Key R$\&$D Program of China (Grant No. 2022YFA1404604), the Chinese Academy of Sciences Project for Young Scientists in Basic Research (Grant No. YSBR-112), the National Natural Science Foundation of China (Grant Nos. 12074400, 62474168, 62293521, 62474168), the Strategic Priority Research Program of the Chinese Academy of Sciences (Grant No. XDB0670303), the Autonomous Deployment Project of the State Key Laboratory of Materials for Integrated Circuits (Grant No. SKLJC-Z2024-B03), and the State Key Laboratory of Advanced Optical Communication Systems and Networks (Grant No. 2024GZKF11).
\end{acknowledgements}

\section*{Author contributions}
J.Z., X.O. J. L. and Y.H. conceived the idea and experiments, J.Z. and X.W. developed the hybrid quantum photonic chip based on III-V and LNOI, X.W. fabricated the hybrid photonic chip and carried out optical measurements. X.W. and J.Z. performed numerical calculations. R. L. and Y. H. grew the QD sample by molecular beam epitaxy. X.W., X.Z., Y.Q. designed and fabricated the GaAs nanophotonic waveguides. X.W. and B. C. fabricated the LNOI photonic chip in collaboration with B.C., Y.Z., J.C., L.D., J.W., and Q.Z.. L.D. and Y.Q. performed wide-field fluorescence imaging of the hybrid quantum photonic chip. J.Z. and X.O. led the project. X. W. and J.Z. wrote the paper with inputs from all the authors.

\section*{Competing interests}
The authors declare no competing interests.
\end{document}